\documentclass[aps,prd,floatfix,showpacs,showkeys,superscriptadress,amsmath,unsortedaddress,nofootinbib,onecolomn]{revtex4-1}
\usepackage{mathtools,amsmath,amssymb,bbm,bm,slashed}
\usepackage{graphicx,float}
\usepackage{hyperref}
\usepackage{color}
\hypersetup{colorlinks=true,linkcolor=blue,citecolor=blue,urlcolor=blue}
\usepackage{morefloats,subfigure}
\newcommand{\sB}{\scriptscriptstyle{(B)}}
\newcommand{\sperp}{\scriptscriptstyle{\perp}}
\newcommand{\shp}{\shortparallel}
\newcommand{\be}{\begin{eqnarray}}
	\newcommand{\ee}{\end{eqnarray}}
\newcommand{\nn}{\nonumber \\}
\newcommand{\sss}[1]{\scriptscriptstyle{#1}}

\def\({\left(}
\def\){\right)}
\makeatletter
\allowdisplaybreaks
\begin{document}
	\title{Neutral pion mass in warm magnetized medium within \textit{linear sigma model coupled to quarks} framework }
	\author{Aritra Das,}
	\email{aritra.das@niser.ac.in} 
	\author{Najmul Haque,}
	\email[]{nhaque@niser.ac.in}
	\affiliation{School of Physical Sciences, National Institute of Science Education and Research,
		An OCC of Homi Bhabha National Institute, Jatni-752050, India}
	
	\begin{abstract}
		We study the neutral pion mass in the presence of an external arbitrary magnetic field in the framework of the linear sigma model coupled to quark (LSMq) at finite temperature. In doing so, we have calculated the pion self-energy, constructed the dispersion equation via re-summation, and solved the dispersion relation at zero three momentum limit. In calculating the pion mass, we have included meson self-coupling's thermal and magnetic contribution and approximate chiral order parameter $v_0$. We report that the $\pi^0$ mass decreases with the magnetic field and increases with temperature.  
	\end{abstract}
	\maketitle
	\section{introduction}
	Recently the properties of hot and dense nuclear matter in the presence of a strong background magnetic field are drawing considerable interest. A transient magnetic field of the order of $10^{18-19}$ Gauss is achieved in the early stage of quark-gluon plasma (QGP) through non-central high-energy heavy-ion collision(HIC)~\cite{Skokov:2009qp, Deng:2012pc,Tuchin:2013apa}. Also, $ B=10^{14}$ Gauss is predicted in the core of neutron stars, and in Magnetars~\cite{Duncan:1992hi}, the primordial magnetic field of $10^{22}$ Gauss could have been present in the early universe due to chiral anomaly~\cite{Joyce:1997uy,Brandenburg:2021aln}. This magnetic field is believed to be responsible for exotic phenomena in the QCD matter in extreme conditions, such as chiral magnetic effect~\cite{Fukushima:2008xe}, magnetic catalysis, inverse magnetic catalysis~\cite{Andersen:2014xxa}, thermal chiral and de-confinement phase transition~\cite{Mizher:2010zb}, superconductivity of QCD vacuum~\cite{Chernodub:2010qx,Chernodub:2011mc}. One of the most important hadrons from the perspective of high energy physics is pion, produced copiously in heavy ion collisions.
	
	The masses of hadrons are expected to be modified under a strong magnetic field. Considering pions as relativistic point particles in the presence of the magnetic field, we expect that $\pi^{\pm}$'s mass increases linearly with $|eB|$ whereas $\pi^0$'s mass remains constant. But the predictions of the properties of pion under the influence of a magnetic field hardly agree with its point-like particle assumption. A recent LQCD study~\cite{Ding:2020hxw,Ding:2020jui}, shows that neutral pion mass decreases with the strength of the magnetic field monotonically. For low $|eB|$, $\pi^{0}$ mass dies rapidly, whereas it saturates with high magnetic field values. This behavior was reproduced to a high degree of accuracy by Ayala et.al.~\cite{Ayala:2020dxs} in the strong field limit $|eB|\gg m_{\pi^0}^2$ using the LSMq model. Also, in ref.~\cite{Das:2019ehv}, authors have reported that neutral pion mass using LSMq model at arbitrary strength of $eB$. By tuning the coupling parameters $\lambda$ and $g$ of the model, they saw that mass decrease from its vacuum value, then found a dip at an intermediate $eB$ and again increases with $eB$. This qualitative non-monotonic behavior is similar to the LQCD study of ref~\cite{Hidaka:2012mz}. Apart from the lattice QCD studies, effective models were also invoked to study meson masses in magnetic backgrounds. For example, in ref~\cite{Adhya:2016ydf}, authors have calculated the magnetic field-dependent pion pole mass considering pseudo-scalar (PS) and pseudo-vector (PV) pion nucleon interaction invoking weak field approximation. They have obtained a decreasing nature for PS coupling while increasing nature for PV coupling for $\pi^0$ mass. Most of the work in determining meson mass under the magnetic field was carried out in NJL model and chiral perturbation theory(ChPt). For example, a full Magnetic Field Independent Regularization (MFIR) scheme with random phase approximation(RPA) method was employed in ref.~\cite{Avancini:2015ady} to calculate meson mass. This MFIR scheme was remarkably~\cite{Avancini:2016fgq} in agreement with LQCD predictions of $\pi^0$ mass. In the ChPt framework, the charged, neutral pion mass was calculated at finite $T$ and $eB$ in ref.~\cite{Andersen:2012zc}. \\ 
	
	In the first part of this series, we calculated neutral pion mass in a magnetic field at zero temperature. We extend the calculation to finite temperature using Linear Sigma Model in this part. It is extensively used to investigate from QCD confinement de-confinement phase transition to properties of hadrons. In the present work, we have examined the behavior of neutral pion mass in the presence of an external magnetic field at a non-zero temperature within the framework of LSMq model. In section~\ref{sec:lsm}, we briefly tour the LSMq model. In section~\ref{sec:self}, we have computed neutral pion self-energy from the LSMq Lagrangian. In section~\ref{sec:pi0_mass}, the $\pi^0$ mass is obtained considering the following three scenarios --- a) bare couplings, b) one-loop corrected meson couplings (Appendix~\ref{sec:vertex}) and c) one-loop corrected meson coupling, as well as quantum corrected effective potential (details of it is elaborated in Appendix~\ref{app:effective_potential}). In section~\ref{sec:results}, the numerical results are discussed, and in section~\ref{sec:conclusion}, we conclude. 
	
	\section{Linear sigma model coupled to quarks}
	\label{sec:lsm}
	The Linear Sigma Model coupled to quark is obtained by appending a $SU(2)$ scalar and pseudoscalar interaction of sigma meson and pion, respectively, with light quarks ($u$ and $d$ flavors). The Lagrangian density for the LSMq reads 
	\begin{align}
		\mathcal{L}&=\frac{1}{2}(\partial_{\mu}\sigma)^2+\frac{1}{2}(\partial_{\mu}\bm{\bm{\pi}})^2+\frac{a^2}{2}(\sigma^2+\bm{\pi}^2)-\frac{\lambda}{4}(\sigma^2+\bm{\pi}^2)^2+i\bar{\psi}\gamma^{\mu}\partial_{\mu}\psi-g\bar{\psi}(\sigma+i\gamma_5\bm{\tau}\cdot\bm{\pi})\psi\ . \label{eq:unb_lag}
	\end{align}
	The first four terms of the last equation is Linear Sigma Model (LSM) part, and the rest two terms are the quark part of $\mathcal{L}$. Here $\bm{\pi}=(\pi^1,\pi^2,\pi^3)$. The physical pion fields are defined as
	\begin{align}
	\pi^{\pm} = \frac{1}{\sqrt{2}}\left(\pi^{1}\pm i \pi^2\right), \qquad \pi^0 = \pi^{3}, 
	\end{align}
	respectively, $\sigma$ is the sigma meson, and  $\psi$ is the light quark doublet with
	\begin{align}
	\psi = \begin{pmatrix}
		u \\ d
	\end{pmatrix},
	\end{align}
	$\bm{\tau} = (\tau^1,\tau^2,\tau^3)$, where $\tau^i$ $(i=1,2,3)$ is the $i^{th}$ Pauli spin matrix.
	Also, $a^2$ is the mass parameter that we take as negative in the symmetry-unbroken state. Finally, $\lambda$ and $g$ are the meson-meson coupling and meson-quark coupling, respectively. The $O(4)$ symmetry of the Lagrangian is spontaneously broken when $a^2$ becomes positive, and the $\sigma$ field gets a non-zero vacuum expectation value (VEV). Hence, after the symmetry breaking, the $\sigma$ field becomes
	\begin{align}
		\sigma \rightarrow \sigma + v.
	\end{align}
	 As a result of this shift, $\mathcal{L}$ reads
	\begin{align}
		\mathcal{L} &= \bar{\psi}(i\gamma^{\mu}\partial_{\mu}-M_{f})\psi+\frac{1}{2}(\partial_{\mu}\sigma)^2+\frac{1}{2}(\partial_{\mu}\mathbb{\bm{\pi}})^2 -\frac{1}{2}M^2_{\sigma}\sigma^2-\frac{1}{2}M^2_{\pi}\bm{\pi}^2-g\bar{\psi}(\sigma+i\gamma_5\bm{\tau}\cdot\bm{\pi})\psi -V(\sigma,\pi)-V_{\rm tree}(v), \label{eq:lag_ssb}
	\end{align}
	with 
	\begin{align}
		V(\sigma,\pi) &= \lambda v \sigma(\sigma^2+\bm{\pi}^2)+\frac{\lambda}{4} (\sigma^2+\bm{\pi}^2)^2, \label{eq:pot_V} \\
		V_{\text{tree}}(v)&=-\frac{1}{2}a^2v^2+\frac{1}{4}\lambda v^4 .\label{eq:pot_tree} 
	\end{align}
	The masses of the quarks, three pions, and sigma are given by 
	\begin{align}
		M_{f} &= gv, \nn 
		M^2_{\pi} &= \lambda v^2 - a^2, \nn
		M^2_{\sigma} &= 3 \lambda v^2 - a^2. \label{eq:tree_masses}
	\end{align}
	respectively. Note that the minimum of the tree-level potential, obtained by solving $\,\,\displaystyle \frac{dV_{\text{tree}}(v)}{dv}\bigg\vert_{v=v_0}=0$, is given by
	\begin{align}
		v_{0} = \sqrt{\frac{a^2}{\lambda}}.
	\end{align}
	Therefore, the masses, evaluated at $v_0$, are given by
	\begin{align}
		M_{f}(v_{0}) &= gv_0, \nn
		M^2_{\pi} &= 0, \nn
		M^{2}_{\sigma} &= 2a^2,
	\end{align}
	after symmetry breaking. To incorporate a non-vanishing pion mass into the model, an explicit symmetry-breaking term is added to the Lagrangian as
	\begin{align}
		\mathcal{L}\rightarrow \mathcal{L}^{\prime} = \mathcal{L}+\mathcal{L}_{ESB} = \mathcal{L} + \frac{1}{2}m_{\pi}^2v (\sigma + v) ,
	\end{align}
	with $m_{\pi} = 0.14 \,\,\text{GeV}$. As a result, the tree-level potential $V_{\text{tree}}$ becomes $V_{\text{tree}}^{\prime} = -\frac{1}{2}(a^2+m^2_{\pi})v^2+\frac{1}{4}\lambda v^4$ and the minimum is shifted to 
	\begin{align}
		v_{0} \rightarrow v^{\prime}_{0} = \left(\frac{a^2+m^2_{\pi}}{\lambda}\right)^{1/2}.
	\end{align}
	The masses, evaluated at this new minimum $v_0^{\prime}$, are given by
	\begin{align}
		M_{f}(v^{\prime}_0) &= g \left(\frac{a^2+m^2_{\pi}}{\lambda}\right)^{1/2}, \label{eq:Mf_mod} \\
		M^2_{\pi}(v^{\prime}_0) &= m^{2}_{\pi}, \nn
		M^2_{\sigma}(v^{\prime}_0) &= 2a^2+3m^2_{\pi}. \label{eq:mass_modified}
	\end{align} 
	The value of $a$ is given by solving Eq.~\eqref{eq:mass_modified} as
	\begin{equation}
		a=\sqrt{\frac{M^2_{\sigma}(v^{\prime}_0)-3M^2_{\pi}(v^{\prime}_0)}{2}}\simeq \sqrt{\frac{m^2_{\sigma}-3m^2_{\pi}}{2}}.
	\end{equation}
	\\ \\
	We consider a homogeneous time-independent background magnetic field in the $z$ direction $\bm{\mathcal{B}} = B \hat{\bm{z}}$, which can be obtained from the electromagnetic four-potential in symmetric gauge $\mathcal{A}^{\mu}=\displaystyle \frac{B}{2}(0,-y,x,0)$. As a result, for the charged d.o.f (quarks and charged pions), the four-derivative $\partial_{\mu}$ is replaced by covariant four-derivative $D_{\mu}=\partial_{\mu}+i\mathcal{Q}\mathcal{A}^{\mu}$. Here, $\mathcal{Q}=q_{f}$ for quarks of flavor $f$ and $\mathcal{Q}=e$ for $\pi^{\pm}$, respectively.

	\section{self-energy of Neutral Pion}\label{sec:self}
	If we rewrite the Feynman diagram of Eq.~\eqref{eq:lag_ssb} in terms of $\pi^{+}$ and $\pi^{-}$ fields, we notice that the neutral pion self-energy $\Pi^{\sB}_0\left(P,T\right)$ has the contributions from $\pi^{\pm}$, $\pi^0$, $\sigma$. It reads
	\begin{align}
		\Pi^{\sB}_0\left(P,T\right) = 8\,\Pi^{\sB}_{\pi^{\pm}}(T) + 12\,\Pi_{\pi^0}(T) + 4\,\Pi_{\sigma}(T) + \sum_{f=u, d}\Pi^{\sB}_{f\bar{f}}(P,T). \label{eq:self_pi0_total}
	\end{align}
	Here, $\Pi^{\sB}_{\pi^{\pm}}(T)$, $\Pi_{\pi^0}(T) $, $\Pi_{\sigma}(T)$ and $\Pi^{\sB}_{f\bar{f}}(P,T)$ are the contribution coming from charged pion, neutral pion, sigma meson, and quark-antiquark loop of flavor, respectively, $f$ to the total $\pi^{0}$ self-energy $\Pi^{\sB}_0\left(P,T\right)$. As mentioned in the previous section, we consider only light flavor in this article which is indicated by the flavor sum over quark-antiquark contribution. Note that the dependence on external momentum $P$ comes in the total self-energy solely from the quark-antiquark part. Also, we have omitted the superscript $B$ from $\pi^0$-loop and $\sigma$-loop contribution since, being charge neutral, they are unaffected by the background magnetic field. In this section, we compute the self-energies indicated on the R.H.S of Eq.~\eqref{eq:self_pi0_total}. Before proceeding, we clarify some notations, conventions, and definitions that will be used repeatedly in the rest of the article. 
	\begin{itemize}
		\item For any generic four vectors $A^{\mu}$, $B^{\mu}$, we adopt the following notation and convention in which four-vectors are denoted by a capital letter ( e.g., $A^{\mu}$ ) and three vectors by small letters with boldface ( e.g., $\bm{a}$) and magnitude by $|\bm{a}|$ or $a$. The following Eq. clearly expresses
		\begin{align}
			&A^{\mu} = (a^0,a^1,a^2,a^3),\qquad A^{\mu}_{\shp} = (a^0,0,0,a^3), \nn &A^{\mu}_{\sperp}=(0,a^1,a^2,0), \qquad (a.b)_{\shp}=a^0b^0-a^3b^3, \nn
			&(a.b)_{\sperp}=a^1b^1+a^2b^2, \qquad A.B = (a.b)_{\shp}-(a.b)_{\sperp},\nn
			&\slashed{a}_{\shp}=(\gamma.a)_{\shp} = \gamma^0a^0-\gamma^3a^3,\,\, \slashed{a}_{\sperp}=(\gamma.a)_{\sperp} = \gamma^1a^1+\gamma^2a^2, \nn
			&a^0=a_0,\quad a_1=-a^1, \quad a_2=-a^2, \quad a_3=-a^3,\nn
			&A^2=a_0^2-a_1^2-a_2^2-a_3^2, \qquad a^2=a_1^2+a_2^2+a_3^2, \nn
			&a_{\shp}^2=a_0^2-a_3^2,\qquad a_{\sperp}^2=a_1^2+a_2^2. \label{eq:notations}
		\end{align}
		\item For calculation involving non-zero temperature, we will work in imaginary time formalism (ITF). In ITF, the integration over the $\text{zero}^{\text{th}}$ component of the four-momentum running in the loop is replaced by a discrete Matsubara frequency sum. We make the following replacement 
		\begin{align}
			\int\limits_{-\infty}^{\infty}\frac{dk_0}{2\pi}\longrightarrow iT\sum_{k_0}\quad,
		\end{align} 
		where
		$k_0 = i\omega_n = i2n\pi T$ for bosons and $k_0=i\tilde{\omega}_n = \mu+i(2n+1)\pi T$ for fermions. Here $T$ and $\mu$ denote the temperature and chemical potential of the thermal medium, respectively. 
		\item The energy and landau level dependent masses in the presence of $B$-field are denoted as follows
		\begin{align}
			M_{\ell, f} = \sqrt{2\ell |q_fB|+M_f^2}, \qquad \Omega_{k,\ell,f} = \sqrt{k_z^2+M_{\ell, f}^2},
		\end{align}	  
		for quarks with flavour $f$, in Landau Level $\ell$ and
		\begin{align}
			m_{\ell,b} = \sqrt{(2\ell + 1) |q_bB|+m_b^2} \qquad E_{k,\ell,b} = \sqrt{k_z^2+m_{\ell, b}^2},
		\end{align}
		for meson with species $b\,(=\pi^{0},\pi^{\pm},\sigma)$. Here we assume $m_u = m_d$ and $m_{\pi^{\pm}} = m_{\pi^0} = m_{\pi}$. Also for charged pions, $q_{\pi^{\pm}} = \pm e$.
		\item In the presence of a magnetic field, the quark propagator takes the following form 
		\begin{align}
			S_f^{\sB}(K) = \exp\left(-\frac{k_{\sperp}^2}{|q_fB|}\right)\sum_{\ell = 0}^{\infty}(-1)^{\ell}\frac{\mathcal{D}_{\ell}(k_{\shp},k_{\sperp},q_fB)}{k_{\shp}^2-2\ell |q_fB|-M^2_{f}+i\varepsilon}, \label{eq:SBf}
		\end{align}
		where
		\begin{align}
			&\mathcal{D}_{\ell}(K,q_{f}B)=4\slashed{k}_{\sperp}L_{\ell-1}^{(1)}\left(\frac{2k_{\sperp}^2}{|q_fB|}\right)\nn
			&\hspace{1cm}+\left(\slashed{k}_{\shp}+M_{f}\right)\left[\left(\mathbbm{1}-\textsf{sgn}(q_{f}B)i\gamma^1\gamma^2\right)L_{\ell}\left(\frac{2k_{\sperp}^2}{|q_fB|}\right)-\left(\mathbbm{1}+\,\textsf{sgn}(q_fB)i\gamma^1\gamma^2\right)L_{\ell-1}\left(\frac{2k_{\sperp}^2}{|q_fB|}\right)\right].
		\end{align}
		Here $L^{(\alpha)}_{\ell}(x)$ is the generalized Laguerre polynomial which is written as
		\begin{align}
			\frac{e^{-\frac{x z}{1-z}}}{(1-z)^{1+\alpha}} = \sum_{\ell=0}^{\infty}L^{(\alpha)}_{\ell}z^n,
		\end{align}
		with $|z|<1$. We note $L_{\ell}^{(0)}(x) = L_{\ell}(x)$ and  $L_{-1}^{(\alpha)} = 0$. Here $\textsf{sgn}$ is the sign function.
		\item 
		In the presence of a magnetic field, the charged boson propagator becomes
		\begin{align}
			D_b^{\sB}(K) = 2\exp\left(-\frac{k_{\sperp}^2}{|eB|}\right)\sum_{\ell = 0}^{\infty}(-1)^{\ell}\frac{L_{\ell}\left(\frac{2k_{\sperp}^2}{|eB|}\right)}{k_{\shp}^2-(2\ell+1)|eB|-m^2_{b}+i\varepsilon}. \label{eq:pion_prop}
		\end{align}
	\end{itemize}
	\subsection{Pion to Quark Anti-quark Loop}
	\label{sec:Piff}
	\begin{figure}[!h]
		\centering
		\includegraphics[scale=0.5]{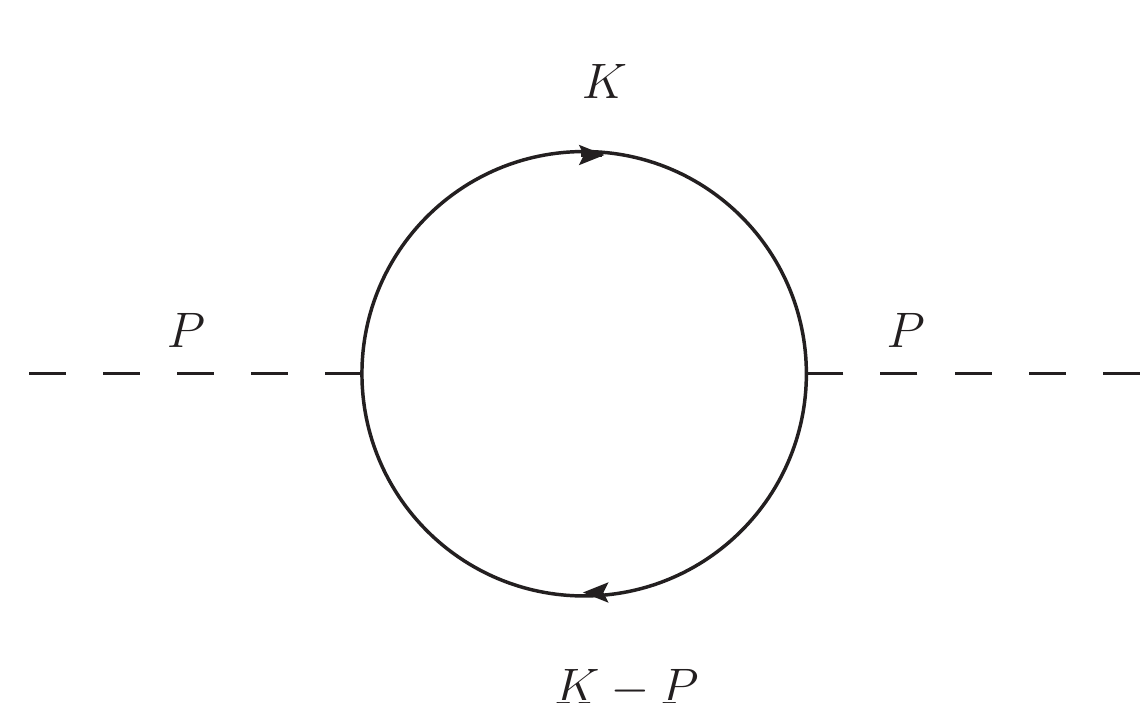}
		\caption{Feynman diagram for one-loop quark anti-quark contribution to the $\pi^{0}$ self-energy}
		\label{fig:pi_quark}
	\end{figure}
	The quark anti-quark contribution to the neutral pion self-energy reads
	\begin{align}
		-i\Pi^{\sB}_{f\bar{f}}(P,T)=N_cg^2\int\frac{d^4K}{(2\pi)^4}\textsf{Tr}\left[\gamma_{5}iS_f^{\sB}(K)\gamma_{5}iS_f^{\sB}(K-P)\right] ,\label{eq:Piff1}
	\end{align}
	where $Q=K-P$. Here $N_c$ denotes the number of colors, which is taken as $3$ for QCD.
	Thus, from Eq.~\eqref{eq:SBf}, the $\Pi^{\sB}_{f\bar{f}}$ in Eq.~\eqref{eq:Piff1} becomes
		\begin{align}
			i\Pi^{\sB}_{f\bar{f}}(P,T) = N_cg^2\int\frac{d^4K}{(2\pi)^4}\exp\left(-\frac{k_{\sperp}^2+q_{\sperp}^2}{|q_fB|}\right)\sum_{\ell,n=0}^{\infty}(-1)^{\ell +n}\frac{\mathcal{N}^{\sB}_{\ell,n}(k_{\shp},k_{\sperp},q_{\shp},q_{\sperp})}{\(k_{\shp}^2-2\ell |q_fB|-M^2_f\)\(q_{\shp}^2-2n |q_{f}B|-M^2_f\)}, \label{eq:Piff2}
		\end{align}
	where 
	\begin{align}
		&\mathcal{N}^{\sB}_{\ell,n}(k_{\shp},k_{\sperp},q_{\shp},q_{\sperp}) = \textsf{Tr}\bigg[\gamma_5\mathcal{D}_{\ell}(k_{\shp},k_{\sperp},q_fB)\gamma_5\mathcal{D}_{n}(q_{\shp},q_{\sperp},q_fB)\bigg].
		\label{eq:num}
	\end{align}
	The trace in \eqref{eq:num} is computed as
		\begin{align}
			&\mathcal{N}^{\sB}_{\ell,n}(k_{\shp},k_{\sperp},q_{\shp},q_{\sperp}) = 8 \left[M_f^2-(k.q)_{\shp}\right] 
			\times\left[ L_{\ell -1}\left(\frac{2k_{\sperp}^2}{|q_fB|}\right)L_{n -1}\left(\frac{2q_{\sperp}^2}{|q_fB|}\right)+L_{\ell}\left(\frac{2k_{\sperp}^2}{|q_fB|}\right)L_{n}\left(\frac{2q_{\sperp}^2}{|q_fB|}\right)\right]\nn
			&\hspace{3cm}+64\left(k^1q^1+k^2q^2\right)L^{1}_{\ell -1}\left(\frac{2k_{\sperp}^2}{|q_fB|}\right)L^{1}_{n -1}\left(\frac{2q_{\sperp}^2}{|q_fB|}\right).
		\end{align}	
	Since we are interested in modification of $\pi^0$ mass, we take the limit $\bm{p}\rightarrow \bm{0}$ of eq.~\eqref{eq:Piff2}. 
	\begin{align}
		\Pi^{\sB}_{f\bar{f}}(p_0,T) = N_cg^2\int\frac{d^4K}{(2\pi)^4}\exp\left(-\frac{2k_{\sperp}^2}{|q_fB|}\right)\sum_{\ell,n=0}^{\infty}(-1)^{\ell +n}\frac{\mathcal{N}^{\sB}_{\ell,n}(k_0,p_0,\bm{q}=\bm{k})}{\left(k_{0}^2-\Omega_{k,\ell,f}^2\right)\left(q_{0}^2-\Omega_{k,n,f}^2\right)}. \label{eq:Piff3}
	\end{align}
	After performing the perpendicular momentum integral analytically, the above expression simplified to
	\begin{align}
		\Pi^{\sB}_{f\bar{f}}(p_0,T)= -N_c8g^2\int\limits_{-\infty}^{\infty}\frac{dk_z}{2\pi}\sum_{\ell,n=0}^{\infty}(-1)^{\ell +n}\frac{\left(\mathcal{I}^{(0)}_{\ell,n,f}+\mathcal{I}^{(0)}_{\ell-1,n-1,f}\right)\big[k_0(k_0-p_0)-k_z^2-M^2_{f}\big]-8\mathcal{I}^{(1)}_{\ell-1,n-1,f}}{\left(k_{0}^2-\Omega_{k,\ell,f}^2\right)\left((k_0-p_0)^2-\Omega_{k,n,f}^2\right)}, \label{eq:Piff5}
	\end{align}
	where $\mathcal{I}^{(\alpha)}_{l,n,f}$ (for $l,n,\alpha\in \mathbb{Z}$ and $l,n,\alpha\geq 0$) is defined in Sec.~\ref{sec:identities}.
	After performing the perpendicular integration according to Eq.~\eqref{eq:perp_intg_result} and employing the following identity (under $k_0$ integral) 
	\begin{align}
		\frac{M_f^2-k_0q_0+k_z^2}{\left(k_0^2-\Omega_{k,\ell,f}^2\right)\left(q_0^2-\Omega_{k,\ell,f}^2\right)}&= \frac{p_0^2-4\ell |q_{f}B|}{2\left(k_0^2-\Omega_{k,\ell,f}^2\right)\left(q_0^2-\Omega_{k,\ell,f}^2\right)}-\frac{1}{k_0^2-\Omega_{k,\ell,f}^2}.
	\end{align} It is convenient to separate the lowest Landau level's and higher Landau level's contributions as follows
	\begin{align}
		\Pi^{\sB}_{f\bar{f},\textsf{LLL}} &= -iN_cg^2\frac{|q_fB|}{\pi}\int\frac{d^2k_{\shp}}{(2\pi)^2}\left[\frac{p_0^2}{2}\frac{1}{\left(k_0^2-\Omega_{k,\ell=0,f}^2\right)\left(q_0^2-\Omega_{k,\ell=0,f}^2\right)}-\frac{1}{k_0^2-\Omega_{k,\ell=0,f}^2}\right], \label{eq:Piff_LLL_vacuum} \\
		\Pi^{\sB}_{f\bar{f},\textsf{HLL}} &= -2N_cig^2\frac{|q_fB|}{\pi}\int\frac{d^2k_{\shp}}{(2\pi)^2}\sum_{\ell =1}^{\infty}\left[\frac{p_0^2}{2}\frac{1}{\left(k_0^2-\Omega_{k,\ell,f}^2\right)\left(q_0^2-\Omega_{k,\ell,f}^2\right)}-\frac{1}{k_0^2-\Omega_{k,\ell,f}^2}\right]. \label{eq:Piff_HLL_vacuum} 
	\end{align}
	In deriving Eq.~\eqref{eq:Piff_HLL_vacuum} and \eqref{eq:Piff_LLL_vacuum}, we used the following identities
	\begin{align}
		\delta_{\ell-1,\ell-1} = 1 - \delta_{0,\ell},\qquad \delta_{\ell,\ell}=1.
	\end{align}
	Here we note that the Kronecker delta gives zero for any negative index.
	This kind of expression is typical in cases involving fermion inside a loop. The degeneracy in higher Landau levels is considered by the factor $(2-\delta_{\ell,0})$. Combining the lowest and higher LL terms we write Eq.~\eqref{eq:Piff_HLL_vacuum} and \eqref{eq:Piff_LLL_vacuum} as
	\begin{align}
		\Pi^{\sB}_{f\bar{f}} (p_0,T) = -ig^2N_c\frac{|q_{f}B|}{\pi}\sum_{\ell =0}^{\infty} \left(2-\delta_{\ell,0}\right)\int\frac{d^2k_{\shp}}{(2\pi)^2}\left[\frac{p_0^2}{2}\frac{1}{\left(k_0^2-\Omega_{k,\ell,f}^2\right)\left(q_0^2-\Omega_{k,\ell,f}^2\right)}-\frac{1}{k_0^2-\Omega_{k,\ell,f}^2}\right]. \label{eq:Piff_Total_vacuum}
	\end{align}	  
	The expression involving higher Landau levels (HLL) has an overall factor of two, which is absent in the expression of the lowest Landau level (LLL). It comes from the fact that the virtual quark-antiquark pair in HLL has spin degeneracy that is lifted in LLL.
	So, after replacing the $k_0$ integration with frequency sum, the expression of $\Pi^{\sB}_{f\bar{f}} (p_0,T) $ becomes 
	\begin{align}
		\Pi^{\sB}_{f\bar{f}} (p_0,T) = N_cg^2\frac{|q_fB|}{\pi}\sum_{\ell =0}^{\infty} \left(2-\delta_{\ell,0}\right)\int_{-\infty}^{\infty}\frac{dk_z}{2\pi}T\sum_{k_0}\left[\frac{p_0^2}{2}\frac{1}{\left(k_0^2-\Omega_{k,\ell,f}^2\right)\left(q_0^2-\Omega_{k,\ell,f}^2\right)}-\frac{1}{k_0^2-\Omega_{k,\ell,f}^2}\right]. \label{eq:Piff_Total_T}
	\end{align}
	In the presence of magnetic field and temperature, any loop integration contains three pieces, i) pure vacuum contribution, ii) pure magnetic field contribution, iii) a thermal as well as a magnetic field (i.e., thermo-magnetic) contribution \footnote{The pure vacuum contribution contains ultraviolet divergence. In our context, the pure magnetic field and thermo-magnetic part are divergence-free. As a result of taking the $B\rightarrow 0$ limit,  the pure magnetic contribution vanishes, and the thermo-magnetic contribution reduces to pure thermal contribution. On the other hand, taking $T\rightarrow 0$ limit, the thermo-magnetic part vanishes }. 
	Now we compute each contribution separately.
	\subsubsection{Pure Vacuum Part}
	For the vacuum part we take $\displaystyle S_{f}(K)=\frac{\slashed{K}+M_f}{K^2-M_f^2}$. The diagram in Fig.~\ref{fig:pi_quark} gives
	\begin{align}
		\Pi^{\textsf{Vacuum}}_{f\bar{f}} (p_0) &= -4N_c i g^2\int\frac{d^4K}{(2\pi)^4} \left[\frac{p_0^2}{2}\frac{1}{k_0^2-M_f^2}\frac{1}{(k_0-p_0)^2-k^2-M_f^2}-\frac{1}{k_0^2-M_f^2}\right].
	\end{align}
	Employing Feynman parametrization technique to the first term in the square bracket, we get
	\begin{align}
		\Pi^{\textsf{Vacuum}}_{f\bar{f}} (p_0) &= -4N_c i g^2\left[\frac{p_0^2}{2}\int_{0}^{1} dx \int\frac{d^4K}{(2\pi)^4}\frac{1}{\left\{(K-xP)^2-\left[M_f^2-x (1-x)\right]\right\}^2} -\int\frac{d^4K}{(2\pi)^4}\frac{1}{K^2-M_f^2}\right].
	\end{align}
	After the change of variable $K-xP\rightarrow K$, we get 
	\begin{align}
		\Pi^{\textsf{Vacuum}}_{f\bar{f}} (p_0) &= -4N_c i g^2\left[\frac{p_0^2}{2}\int_{0}^{1} dx \int\frac{d^4K}{(2\pi)^4}\frac{1}{\left\{K^2-\left[M_f^2-x (1-x)\right]\right\}^2} -\int\frac{d^4K}{(2\pi)^4}\frac{1}{K^2-M_f^2}\right].
	\end{align}
	The above integral is ultra-violet divergent that is regularized by the method of dimensional regularization. We analytically continue the momentum integration to $d$ dimensions. Now, it can be performed by using the following identity
	\begin{align}
		\int\frac{d^dK}{(2\pi)^d}\frac{1}{\left(K^2-\Delta\right)^{\alpha}} = i\frac{\pi^{d/2}}{(2\pi)^d}\frac{(-1)^{\alpha}}{\Gamma(\alpha)}\frac{\Gamma\left(\alpha-\frac{d}{2}\right)}{\Delta^{\alpha-\frac{d}{2}}}.
	\end{align}	 
	It yields, after taking out a auxiliary scale factor $\Lambda$ from $g$ as $g\rightarrow g\Lambda^{2-\frac{d}{2}}$, to   
	\begin{align}
		\Pi^{\textsf{Vacuum}}_{f\bar{f}} (p_0) &= 4N_cg^2 \Lambda^{4-d}\frac{1}{(4\pi)^{d/2}}\left\{\frac{p_0^2}{2}\int_{0}^{1}dx\frac{\Gamma\left(2-\frac{d}{2}\right)}{\left[M_f^2-x (1-x) p^2_0\right]^{2-\frac{d}{2}}}+\frac{\Gamma\left(1-\frac{d}{2}\right)}{\left(M_f^2\right)^{1-\frac{d}{2}}}\right\}.
	\end{align}        
	Lets us take $\epsilon =2-\dfrac{d}{2}$ and obtain
	\begin{align}
		\Pi^{\textsf{Vacuum}}_{f\bar{f}} (p_0) &= N_c\frac{g^2}{4\pi^2}\left[\frac{p_0^2}{2}\int\limits^{1}_{0}dx\,\,\left(\frac{1}{4\pi\Lambda^2}\right)^{-\epsilon}\frac{\Gamma(\epsilon)}{\left[M_f^2-x (1-x) p^2_0\right]^{\epsilon}}+\left(\frac{1}{4\pi\Lambda^2}\right)^{-\epsilon}\frac{\Gamma(\epsilon-1)}{\left(M_f^2\right)^{\epsilon-1}}\right].
	\end{align}
	Now, we expand the above expression around $\epsilon = 0$ to get
	\begin{align}
		\Pi^{\textsf{Vacuum}}_{f\bar{f}} (p_0) &=N_c\frac{g^2}{4\pi^2}\bBigg@{3}[ \left(\frac{p_0^2}{2}-M_f^2\right)\left(\frac{1}{\epsilon}-\gamma_{E}+\log(4\pi\Lambda^2)\right)\nn
		&\hspace{2cm}-\left\{\frac{p_0^2}{2}\int\limits_{0}^{1} dx\, \log\Big[M_f^2-x(1-x)p_0^2\Big]+M_f^2-M_f^2\log(M_f^2)\right\}\bBigg@{3}].
	\end{align}
	In accordance with $\overline{\text{MS}}$, prescription we absorb the $\dfrac{1}{\epsilon}-\gamma_{E}+\log(4\pi)$ by introducing counter-term. It leads to 
	\begin{align}
		\Pi^{\textsf{Vacuum}}_{f\bar{f},\overline{\text{MS}}} (p_0) &=N_c\frac{g^2}{4\pi^2}\left[ \frac{p_0^2}{2}\int_{0}^{1} dx \log\left(\frac{\Lambda^2}{M_f^2-x(1-x)p_0^2}\right)-M_f^2\left(\log\frac{\Lambda^2}{M_f^2}+1\right)\right].
	\end{align}
	\subsubsection{Magnetic field part}
	After getting the pure vacuum part, we now evaluate the magnetic part. We take $\epsilon = 1-\dfrac{d}{2}$ so that the integral formally diverges at $\epsilon = 0$. We get
	\begin{align}
		\Pi^{\sB}_{f\bar{f}}(p_0) &= N_c\frac{g^2}{4\pi^2}|q_fB| \sum_{\ell = 0}^{\infty} \left(2-\delta_{\ell, 0}\right)\left[\frac{p_0^2}{2}\int_{0}^{1} dx \frac{1}{(4\pi \Lambda^2)^{-\epsilon}}\frac{\Gamma (1+\epsilon )}{\left[M_f^2+2\ell |q_{f}B|-x (1-x) p_0^2\right]^{1+\epsilon}}\right.\nn
		&\left.\hspace{5cm}+\ \ \frac{1}{(4\pi \Lambda^2)^{-\epsilon}}\frac{\Gamma (\epsilon )}{\left(M_f^2+2\ell |q_{f}B|\right)^{\epsilon}}\right].
	\end{align}
	The sum over the Landau levels in the R.H.S of the last equation can be performed as
	\begin{align}
		&\sum_{\ell = 0}^{\infty} \frac{2-\delta_{\ell,0}}{\left[M_f^2+2 \ell |q_fB|- x (1-x) p^2_0\right]^{1+\epsilon}} = \frac{(2|q_fB|)^{-\epsilon}}{|q_fB|}\zeta\left(1+\epsilon, \frac{M_f^2-x (1-x) p_0^2}{2|q_fB|} \right) - \frac{1}{\left(M_f^2-x (1-x) p_0^2\right)^{1+\epsilon}}, \\
		&\sum_{\ell = 0}^{\infty}\frac{2-\delta_{\ell,0}}{\left(M_f^2+2\ell |q_fB|\right)^{\epsilon}} = 2 (2|q_fB|)^{-\epsilon} \zeta\left(\epsilon, \frac{M_f^2}{2|q_fB|}\right)-\frac{1}{M_f^{2\epsilon}}.
	\end{align}
	Thus, we have 
	\begin{align}
		\Pi^{\sB}_{f\bar{f}}(p_0) &= N_c\frac{g^2}{4\pi^2}|q_fB|\bBigg@{3}\{ \frac{p_0^2}{2}\frac{\Gamma (1+\epsilon )}{(4\pi\Lambda^2)^{-\epsilon}}\int_{0}^{1}dx\left[\frac{(2|q_fB|)^{-\epsilon}}{|q_fB|}\zeta\left(1+\epsilon,\frac{M_{f}^2-x (1-x)p_0^2}{2|q_fB|}\right)\right.\nn
		&\left.\left.-\frac{1}{\left(M_f^2-x (1-x) p_0^2\right)^{1+\epsilon}}\right] + \frac{\Gamma (\epsilon)}{(4\pi\Lambda^2)^{-\epsilon}}\bBigg@{3}[2 (2 |q_fB|)^{-\epsilon}\zeta\left(\epsilon,\frac{M_f^2}{2 |q_fB|}\right)-\frac{1}{M_f^{2\epsilon}}\bBigg@{3}]\right\}.
	\end{align}
	As usual if we expand the above expression around $\epsilon = 0$, we get terms of the form $\zeta\left(0,x\right)$ and $\partial_s\zeta (s,x)|_{s=0}$. Now, using the following properties of the Hurwitz zeta function 
	\begin{align}
		\zeta(0,x)=\frac{1}{2}-x, \quad \zeta^{(1,0)}(0,x)\equiv \frac{d}{ds}\zeta(s,x)\big\vert_{s=0} = \log \Gamma (x) -\frac{1}{2}\log (2\pi)
	\end{align} 
	and after performing some simplifications, we obtain
	\begin{align}
		\Pi^{\sB}_{f\bar{f}}(p_0) =&N_c\frac{g^2}{4\pi^2}\Bigg\{\left(\frac{p_0^2}{2}-M_f^2\right)\left[\frac{1}{\epsilon}-\gamma_{E}+\log(4\pi\Lambda^2)\right]+|q_fB|\left[ 2\log\Gamma\left(\frac{M_f^2}{2|q_fB|}\right)+\log\left(\frac{M_f^2}{4\pi|q_fB|}\right)\right]\nn
		&\hspace{1cm}-\left(\frac{p_0^2}{2}-M_f^2\right)\log(2|q_fB|)-\frac{p_0^2}{2}\int_{0}^{1}dx\,\,\left[\psi\left(\frac{M_f^2-x (1-x)p_0^2}{2 |q_fB|}\right) + \frac{|q_fB|}{M_f^2-x (1-x)p_0^2}\right]\Bigg\}.
	\end{align}	  
	After performing the integral, we get
	\begin{align}
		\int_{0}^{1}dx\,\,\frac{|q_fB|}{M_f^2-x (1-x)p_0^2} = \frac{4|q_fB|}{p_0^2}\cot^{-1}\left(\sqrt{\frac{4 M_f^2}{p_0^2}-1}\right),
	\end{align}
	the (\ Vacuum + Magnetic field dependent )\ part of self-energy as
	\begin{align}
		\Pi^{\sB}_{f\bar{f}}(p_0) =&N_c\frac{g^2}{4\pi^2}\left(\frac{p_0^2}{2}-M_f^2\right)\left(\frac{1}{\epsilon}-\gamma_{E}+\log(4\pi\Lambda^2)-\log(2|q_fB|)\right)-|q_fB|\left[ 2\log\Gamma\left(\frac{M_f^2}{2|q_fB|}\right)+\log\left(\frac{M_f^2}{4\pi|q_fB|}\right)\right.\nn
		&\left.\hspace{4cm}+2\cot^{-1}\left(\sqrt{\frac{4 M_f^2}{p_0^2}-1}\right)\right]-\frac{p_0^2}{2}\int_{0}^{1}dx\,\,\psi\left(\frac{M_f^2-x (1-x)p_0^2}{2 |q_fB|}\right). \label{eq:vacuum}
	\end{align}	
	After subtracting the vacuum part from Eq.~\eqref{eq:vacuum}, we get pure magnetic field-dependent contribution as
		\begin{align}
			&\Pi^{\sB}_{f\bar{f}}(p_0) - \Pi^{\textsf{Vacuum}}_{f\bar{f}}(p_0) = N_c\frac{g^2}{4\pi^2}\bBigg@{3}\{\frac{p_0^2}{2}\int_{0}^{1}dx\left[\log\frac{M_f^2-x (1-x)p_0^2}{2|q_fB|}-\psi\left(\frac{M_f^2-x (1-x)p_0^2}{2|q_fB|}\right)-\frac{|q_fB|}{M_f^2-x (1-x)p_0^2}\right]\nn
			&\hspace{1cm}-2|q_fB|\left[\log\Gamma\left(\frac{M_f^2}{2|q_fB|}\right)+\log\left(\frac{M_f^2}{4\pi|q_fB|}\right)\right]+M_f^2-M_{f}^2\log\left(\frac{M_{f}^2}{2|q_{f}B|}\right)\bBigg@{3}\}.
		\end{align}
	\subsubsection{The thermomagnetic part}
	To get the thermo-magnetic part, we need to perform the fermionic frequency sums. It is performed in Appendix~\ref{sec:freq_sums_fermionic}. Here we quote the results
	\begin{align}
		&T \sum_{k_0}\frac{1}{k_0^2-\Omega_{\ell,k,f}^2}\frac{1}{(k_0-p_0)^2-\Omega_{\ell,k,f}^2} = -\frac{1-\widetilde{n}^{+}(\Omega_{\ell,k,f})-\widetilde{n}^{-}(\Omega_{\ell, k})}{\Omega_{\ell, k,f}\left(p_0^2-4\Omega_{\ell, k,f}^2\right)},\nn
		&T \sum_{k_0}\frac{1}{k_0^2-\Omega_{\ell,k,f}^2}=-\frac{1-\widetilde{n}^{+}(\Omega_{\ell,k,f})-\widetilde{n}^{-}(\Omega_{\ell, k,f})}{2 \Omega_{\ell, k,f}}.
	\end{align}
	Now, substituting the frequency sums in the last line in Eq.~\eqref{eq:Piff_Total_T} and simplifying, we arrive
	\begin{align}
		\Pi^{\sB}_{f\bar{f}}(p_0,T) = -N_c\frac{g^2}{2\pi^2}|q_fB|\sum_{\ell = 0}^{\infty}\left(2-\delta_{\ell,0}\right)\int_{-\infty}^{\infty}dk_{z}\,\Omega_{\ell, k,f}\frac{1-\widetilde{n}^{+}(\Omega_{\ell,k,f})-\widetilde{n}^{-}(\Omega_{\ell, k,f})}{p_0^2-4\Omega_{\ell, k,f}^2}.
	\end{align}
	We have calculated the vacuum + pure $B$ part earlier, which comes from $1$ with the distribution function $\widetilde{n}^{\pm}$. So dropping that term, we get the thermo-magnetic part as
		\begin{align}
			\Pi^{\sB}_{f\bar{f},\textsf{ThM}}(p_0,T) = N_c\frac{g^2}{2\pi^2}|q_fB|\sum_{\ell = 0}^{\infty}\left(2-\delta_{\ell,0}\right)\int_{-\infty}^{\infty}dk_z\,\Omega_{\ell, k,f}\frac{\widetilde{n}^{+}(\Omega_{\ell,k,f})+\widetilde{n}^{-}(\Omega_{\ell, k,f})}{p_0^2-4\Omega_{\ell, k,f}^2}.
		\end{align}
	\subsection{Pion to pion loop}
	\subsubsection{Charged Pion Contribution}
	\begin{figure}[!h]
		\centering
		\includegraphics[scale=0.5]{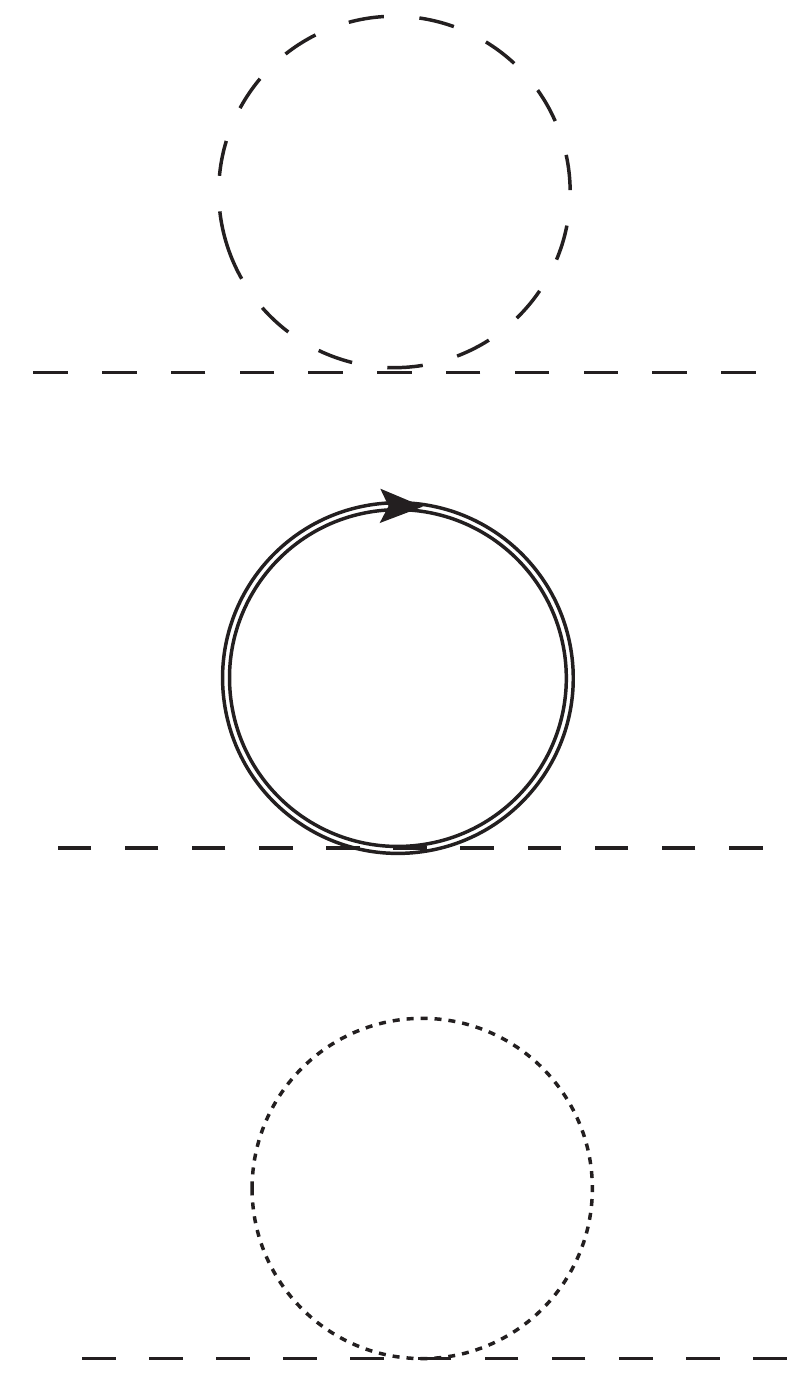}
		\caption{Feynman diagram for one-loop charged pion, neutral pion, and sigma meson contribution to the $\pi^{0}$ self-energy. Only the charged pion is affected by the magnetic field and temperature, but the neutral pion and the sigma meson are affected by only temperature. }
		\label{fig:pi_pi}
	\end{figure}
	The tadpole diagram reads 
	\begin{align}
		\Pi^{\sB}_{\scriptscriptstyle{\pi^{\pm}}} = \frac{\lambda}{4}\int\dfrac{d^4K}{(2\pi)^4}i D^{\sB}(K). \label{eq:Pipi1}
	\end{align}
	Substituting Eq.~\eqref{eq:pion_prop} in Eq.~\eqref{eq:Pipi1}, we get
	\begin{align}
		&\Pi^{\sB}_{\scriptscriptstyle{\pi^{\pm}}} = i\frac{\lambda}{2}\int\dfrac{d^2k_{\shp}}{(2\pi)^2}\sum_{\ell =0}^{\infty}\frac{\mathcal{J}_{\ell}}{k_{\shp}^2-(2\ell+1)|eB|-m^2_{\pi}+i\varepsilon}, \label{eq:Pipi2}
	\end{align}
	where we have defined
	\begin{align}
		\mathcal{J}_{\ell} = \int\dfrac{d^2k_{\sperp}}{(2\pi)^2}(-1)^{\ell}\exp\left(-\frac{k_{\sperp}^2}{|eB|}\right)L_{\ell}\left(\frac{2k_{\sperp}^2}{|eB|}\right). \label{J_int}
	\end{align}
	Here the integral can be performed analytically and shown in Appendix~\ref{sec:freq_sums}. Here is the result quoted:
	\begin{align}
		\mathcal{J}_{\ell} = \dfrac{|eB|}{4\pi}.
	\end{align} 
	This leads Eq.~\eqref{eq:Pipi2} to
	\begin{align}
		&\Pi^{\sB}_{\scriptscriptstyle{\pi^{\pm}}} =i\dfrac{\lambda}{2}\dfrac{|eB|}{4\pi}\sum_{\ell =0}^{\infty}\int\dfrac{d^2k_{\shp}}{(2\pi)^2}\frac{1}{k_{\shp}^2-m_{\ell,\pi}^2}. \label{eq:Pipi3}
	\end{align}
	The integral in Eq.~\eqref{eq:Pipi3} is divergent. To regulate the divergence, we go to $d = 2 - 2\epsilon$ dimension. Also, from the dimensional argument, we take out a dimensional quantity via an auxiliary scale by replacing $\lambda\rightarrow \Lambda^{d-2}\lambda$. So the integral becomes
	\begin{align}
		&\Pi^{\sB}_{\scriptscriptstyle{\pi^{\pm}}} =i\dfrac{\lambda\Lambda^{2-d}}{2}\dfrac{|eB|}{4\pi}\sum_{\ell =0}^{\infty}\int\dfrac{d^dk_{\shp}}{(2\pi)^d}\frac{1}{k_{\shp}^2-m_{\ell,\pi}^2}. \label{eq:Pipi_magnetic}
	\end{align}
	After performing the $d-$dimensional integral~\cite{Das:2008zze}, we get
	\begin{align}
		&\Pi^{\sB}_{\scriptscriptstyle{\pi^{\pm}}} =-\lambda\Lambda^{2-d}\dfrac{|eB|}{8\pi}\sum_{\ell =0}^{\infty}\frac{\pi^{d/2}}{(2\pi)^{d/2}}\frac{\Gamma\left(1-\frac{d}{2}\right)}{\left(m^2_{\ell,\pi}\right)^{1-\frac{d}{2}}}.
	\end{align}
	Now we write the last equation in terms of $\epsilon$ to get
	\begin{align}
		&\Pi^{\sB}_{\scriptscriptstyle{\pi^{\pm}}} = \frac{|eB|}{32\pi^2}\left(\frac{2|eB|}{4\pi\Lambda^2}\right)^{-\epsilon}\Gamma(\epsilon)\sum_{\ell=0}^{\infty}\frac{1}{\left(\ell+\frac{1}{2}+\frac{m_{\pi}^2}{2|eB|}\right)^{\epsilon}}.
	\end{align}
	Now we expand the last equation around $\epsilon = 0$ to get
	\begin{align}
		&\Pi^{\sB}_{\scriptscriptstyle{\pi^{\pm}}} = \frac{\lambda|eB|}{32\pi^2}\left\{\zeta\left(0,\frac{1}{2}+\frac{m_{\pi}^2}{2|eB|}\right)\left[\frac{1}{\epsilon}-\gamma_{E}+\log(4\pi)-\log\frac{2|eB|}{\Lambda^2}\right]+\zeta^{(1,0)}\left(0,\frac{1}{2}+\frac{m_{\pi}^2}{2|eB|}\right)\right\} \nn
		&=-\frac{\lambda m_{\pi}^2}{64\pi^2}\left[\frac{1}{\epsilon}-\gamma_E+\log 4\pi +\log \frac{\Lambda^2}{2|eB|}+\frac{|eB|}{m^2_{\pi}}\log 2\pi-\frac{2|eB|}{m^2_{\pi}}\log\Gamma\left(\frac{1}{2}+\frac{m_{\pi}^2}{2|eB|}\right)\right].
	\end{align}
	We can get weak field results by using the asymptotic expansion~\cite{Asymptotic_Elezovic}
	\begin{align}
		\log\Gamma\left(t+x\right) \sim \left(t+x-\frac{1}{2}\right)\log x-x+\frac{1}{2}\log(2\pi)+\sum_{n=1}^{\infty}(-1)^n\frac{1}{n(n+1)}B_{n}(t)\frac{1}{x^n},
	\end{align}
	where $B_{n}(t)$ are the Bernoulli polynomials defined by generating function
	\begin{align}
		\frac{e^{x\,t}}{e^t-1} = \sum_{n=0}^{\infty}\frac{B_n(t)}{n!}x^{n-1}.
	\end{align}
	Starting from Eq.~\eqref{eq:Pipi3}, we obtain
	\begin{align}
		\Pi^{\sB}_{\scriptscriptstyle{\pi^{\pm}}} = -\dfrac{\lambda}{2}\dfrac{|eB|}{4\pi}\sum_{\ell =0}^{\infty}\,\,\int\limits_{-\infty}^{\infty}\dfrac{dk^{3}}{2\pi}\,\,T\!\sum_{k_0}\frac{1}{k_{0}^2-E_{k,\ell}^2}. \label{eq:PipiTM_1}
	\end{align}
	Here the frequency sum is performed in Appendix~\ref{sec:freq_sums}, and the result is
	\begin{align}
		T\sum_{k_0}\frac{1}{k_{0}^2-E_{k,\ell}^2} = -\frac{1+2n\left(E_{k,\ell}\right)}{2 E_{k,\ell}}, \label{eq:freq_sum_b_result}
	\end{align}
	where 
	\begin{align}
		n(E_{k,\ell}) = \dfrac{1}{\exp\left(E_{k,\ell}/T\right)-1}.
	\end{align}
	Thus after plugging Eq.~\eqref{eq:freq_sum_b_result} in Eq.~\eqref{eq:PipiTM_1} and dropping the non-thermal term (containing $1$ in frequency sum), we get the thermo-magnetic contribution as 
	\begin{align}
		\Pi^{\sB}_{\scriptscriptstyle{\pi^{\pm}}} = \frac{\lambda |eB|}{16 \pi^2}\sum_{\ell=0}^{\infty} \int\limits_{-\infty}^{\infty} dk_z \,\,\frac{n\left(E_{k,\ell}\right)}{E_{k,\ell}}.
	\end{align}
	\subsubsection{Neutral Pion and Sigma loop Contribution}
	For the neutral pion and sigma loop, there will be no effects from the magnetic field as they are chargeless. Again we drop the vacuum part and consider only the thermal correction. 
	\begin{align}
		\Pi_{j} = i\frac{\lambda}{4}\int\frac{d^4K}{(2\pi)^4}D_{j}(K), 
	\end{align}
	where $D_{j}(K) = \dfrac{1}{K^2-m_j^2}$ is the propagator for the $j$-type particle with $j=\pi^{0}, \sigma$. After doing the usual replacements and performing the frequency sum, we arrive at
	\begin{align}
		\Pi^{\text{Th}}_{j} = \frac{\lambda}{8\pi^2}\int\limits_{0}^{\infty} dk \,k^2 \frac{n\left(\sqrt{k^2+m_j^2}\right)}{\sqrt{k^2+m_j^2}}.
	\end{align}
	\section{$\pi^0$ mass} \label{sec:pi0_mass}
	In this section, we compute neutral pion mass. We need to solve the following equation
	\begin{align}
		p_0^2 - |\bm{p}|^2 -m_{\pi}^2-\Pi^{\sB}\left(p_0,\bm{p},T\right) = 0 \label{eq:basic_disp}
	\end{align}	
	in the limit of $\bm{p}\rightarrow \bm{0}$ and $p_0 = M_{\pi}\left(|eB|,T\right)$. The self-energy is given by~\eqref{eq:self_pi0_total}. Note that there is a factor of $g^2$ in the expression of quark loop contribution as indicated in \eqref{eq:Piff1} and a factor of $\lambda$ in the expression of meson loop contribution. We shall solve Eq.~\eqref{eq:basic_disp} in three settings as follows
	\subsection{Basic Case}
	In this case, we take a numerical value of $\lambda$, $g$, which are two coupling parameters of the theory. Also, we consider the vacuum value of $m_{\pi} = 0.14$ GeV. Then we solve Eq.~\eqref{eq:basic_disp}.
	\subsection{Including Self-Coupling}
	Here we consider one-loop correction of vertex $\lambda$. in this case we take one-loop modified vertex $\lambda_{\textsf{eff}} = \lambda + \Delta\lambda$, where $\Delta\lambda$ is given by Eq.~\eqref{eq:delta_lambda}. So the dispersion relation becomes
	\begin{align}
		p_0^2 -(\lambda_{\textsf{eff}} (v_{0}^{\prime})^2 - a^2)-\Pi^{\sB}_{\text{VM}}\left(p_0,\bm{p}\rightarrow\bm{0},T\right) = 0. \label{eq:eff_vertex_disp}
	\end{align}	  
	Here the VM subscript in $\Pi$ indicates that we replace the expression of $\lambda_{\text{eff}}$ in place of $\lambda$ that appears in front of meson self-energy contribution. As mentioned earlier, $v^{\prime}_{0}$ is the minimum of the tree-level potential.
	\subsection{Including Self-Coupling and Quantum Corrected Minimum of Effective Potential}
	Here we solve Eq.~\eqref{eq:eff_vertex_disp} with $v_{0}^{\prime}$ substituted by $v_{B}(T)$, namely
	\begin{align}
		p_0^2 -(\lambda_{\textsf{eff}}\,\,v_{B}^2(T) - a^2)-\Pi^{\sB}_{\text{VM}}\left(p_0,\bm{p}\rightarrow\bm{0},T\right) = 0, \label{eq:eff_vertex_disp_T}
	\end{align}
	where $v_{B}(T)$ is the minimum of the effective potential. The topic of effective potential is discussed in detail in  Appendix~\ref{app:effective_potential}.
	\section{Results}\label{sec:results}
	We have plotted the magnetic field and temperature dependence of neutral pion mass in a warm magnetized medium. The Lagrangian has the following parameters --- the boson self coupling $\lambda$, boson-fermion coupling $g$, vacuum pion mass $m_{\pi}$, mass parameter $a$.\footnote{In our case $a$ is fixed by $m_{\pi}$ and $m_{\sigma}$. So we can think of $m_{\sigma}$ as a parameter of the theory instead of $a$.} We have kept the value of $m_{\pi}$ at $0.14$ GeV and $m_{\sigma}$ at $0.435$ GeV throughout. We have kept the temperature in all of our plots up to $140$ MeV, which is less than or equal to the chiral phase transition temperature (in LQCD, it is calculated to $\sim\,156$ MeV~\cite{Bali:2012zg}). For the magnetic field, we considered it up to $20m_{\pi}^2 = 0.392$ GeV, which is beyond the magnitude generated in heavy ion collision inside the core of Magnetars. We have taken $\lambda = 3.67$ and $g=0.46$, which was used by Ayala et al. in Ref~\cite{Ayala:2020dxs} to match their result of the magnetic field dependence of $\pi^0$ mass, calculated in strong magnetic field approximation at zero temperature, with the LQCD data of Ref.~\cite{Ding:2020hxw}. In our calculation, we have tackled the sum over Landau levels and integration over $k_z$ appearing in the thermo-magnetic part of self-energy as well as effective potential numerically. For a very low magnetic field, one can note that the result saturates by summing over $\sim 5,000$ LL's. In our calculation, we have taken $\ell_{\textsf{Max}} = 10,000$, i.e., we summed over $10,000$ Landau levels, which is more than enough to reach saturation.      \\  
	\begin{figure}
		\centering
		\includegraphics[scale=0.4]{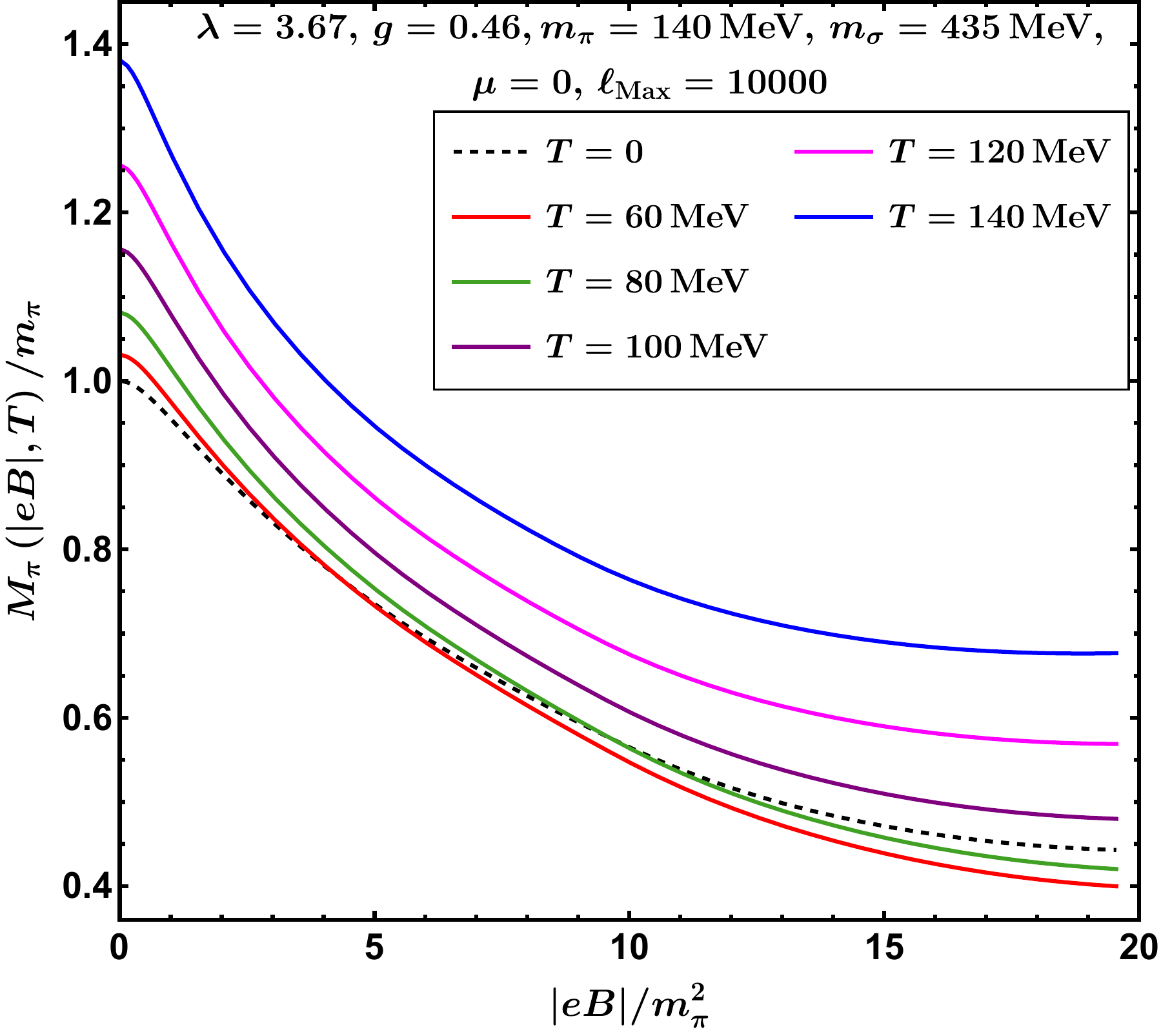}
		\includegraphics[scale=0.4]{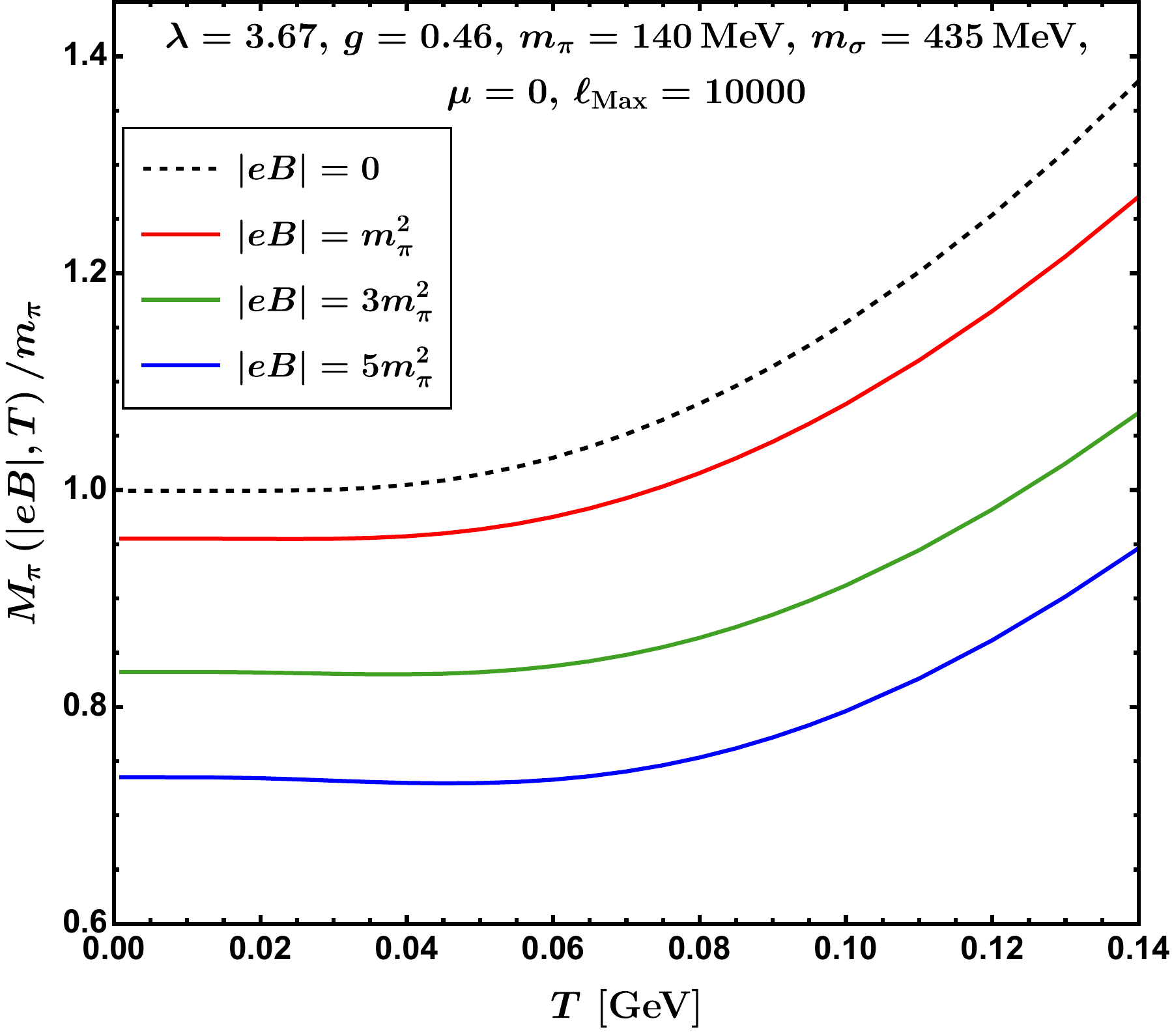}
		\label{fig:full_mpi2_T_eB}
		\caption{[color online] Figure shows the behaviour of neutral pion $\pi^0$ mass. The left panel shows the variation of $\pi^0$ mass with background magnetic field $|eB|$ at some of the fixed values of temperature ranging from $0$ to $140$ MeV. The right panel shows the plot of $\pi^0$ mass with temperature keeping the value of the magnetic field fixed. The $X$-axis and $Y$ are scaled with the square of vacuum pion mass $m^2_{\pi^0}$ and the vacuum pion mass $m_{\pi^0}$. In this plot, we have taken the  minimum of effective potential $v_{B}(T)$ and one-loop meson self-coupling $\lambda_{\textsf{eff}}$ to obtain pion mass as indicated by Eq.~\eqref{eq:eff_vertex_disp_T}  }
	\end{figure}
	In Fig.~\ref{fig:full_mpi2_T_eB}, we show the plots of neutral pion mass with $|eB|$ (left panel) and with $T$ (right panel) considering the effect of effective vertex and quantum corrected condensate $v_B(T)$. The mass decreases with increasing $|eB|$. The fall is rapid at low values of $|eB|$ for all temperate. But after a certain value ($~15m_{\pi}^2$), it saturates with the field. Note that as we increase temperature, the fall with $|eB|$ becomes more rapid in the temperature range $\sim (0-100)$ MeV within the window of $|eB|~ 5m_{\pi}^2-15m_{\pi}^2$. As a result, the plot in the right panel of Fig.~\ref{fig:full_mpi2_T_eB} $T=0$ intersects with $T=60$ MeV and $T=80$ MeV. Now for the variation with temperature, the mass increases with temperature, which is quite expected as increasing temperature gives more thermal contribution. The mass remains uniform for low $T$ but sharply increases after $60$ MeV. This behavior is observed for all values of magnetic fields considered, and it can be explained from the plot of mass with the magnetic field.   At low temperatures, for the magnetic field values considered in the right panel of Fig~\ref{fig:full_mpi2_T_eB}, the field has a much stronger tendency to suppress the mass than the temperature to enhance it. But as the temperature increases, it gradually becomes more dominating than the magnetic field. \\ \\
	\begin{figure}
		\centering
		\includegraphics[scale=0.4]{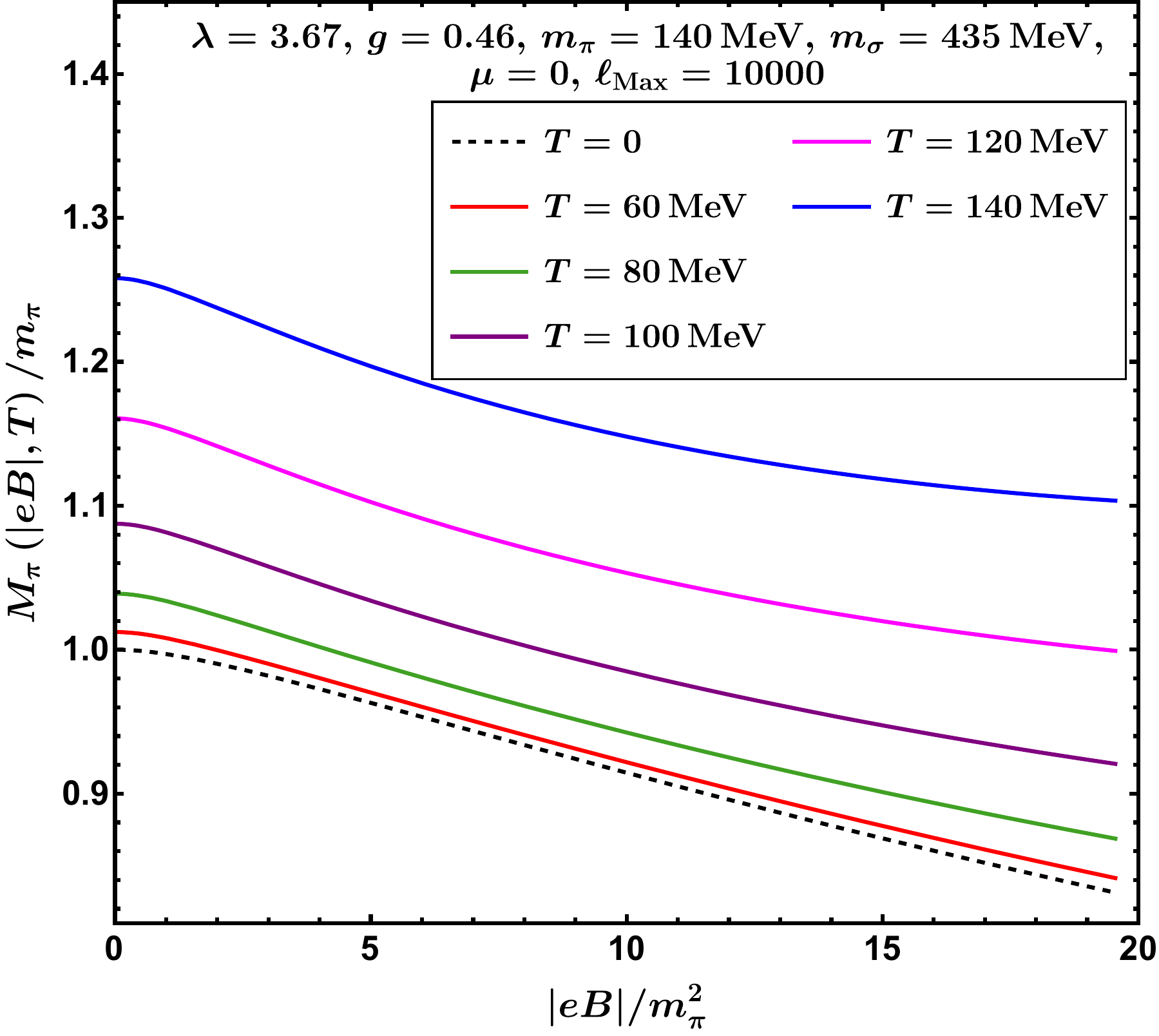}
		\includegraphics[scale=0.4]{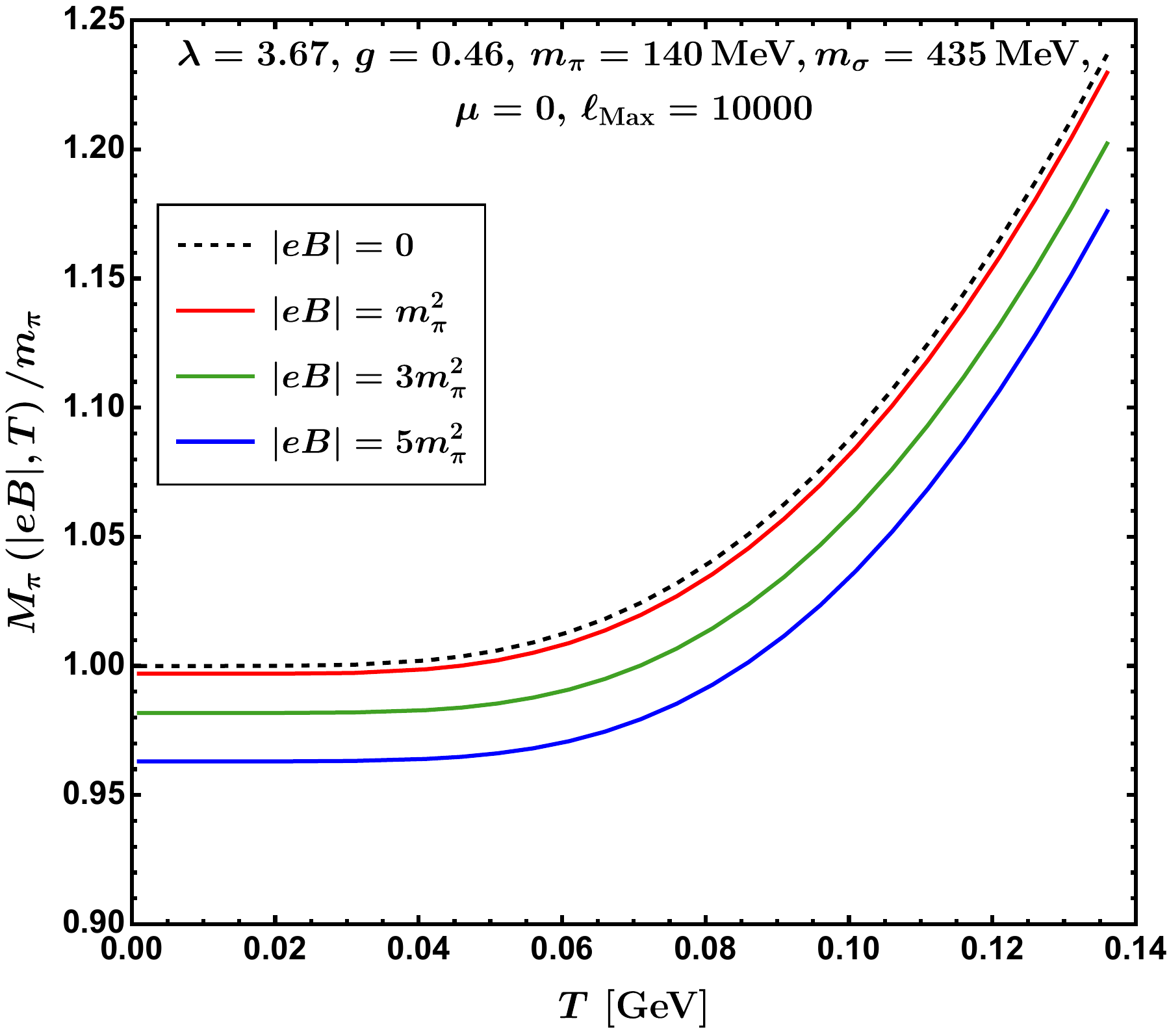}
		\label{fig:basic_mpi2_T_eB}
		\caption{[color online] Figure shows the behaviour of neutral pion $\pi^0$ mass. The left panel shows the variation of $\pi^0$ mass with background magnetic field $|eB|$ at some of the fixed values of temperature ranging from $0$ to $140$ MeV. The right panel shows the plot of $\pi^0$ mass with temperature keeping the value of the magnetic field fixed. The $X$-axis and $Y$ are scaled with the square of vacuum pion mass $m^2_{\pi^0}$ and the vacuum pion mass $m_{\pi^0}$. In this plot, we have taken the classical minimum $v_0^{\prime}$ and bare meson self coupling $\lambda$ in order to obtain pion mass as indicated by Eq.~\eqref{eq:basic_disp}  }
	\end{figure}
	To compare the effect of including effective meson vertex and $V_B(T)$ with Fig.~\ref{fig:full_mpi2_T_eB}, we have added the variation of neutral pion mass with $|eB|$ and $T$ in Fig.\ref{fig:basic_mpi2_T_eB} for the very naive case in which we considered the classical $v_0^{\prime}$ and bare meson-meson coupling $\lambda$. As we can see clearly, the mass falls somewhat less steeply than Fig.~\ref{fig:full_mpi2_T_eB} with the magnetic field. But the rise of mass with temperature is more pronounced and steeper than that with including effective vertex and $v_B(T)$.  
	\begin{figure}
		\centering
		\includegraphics[scale=0.4]{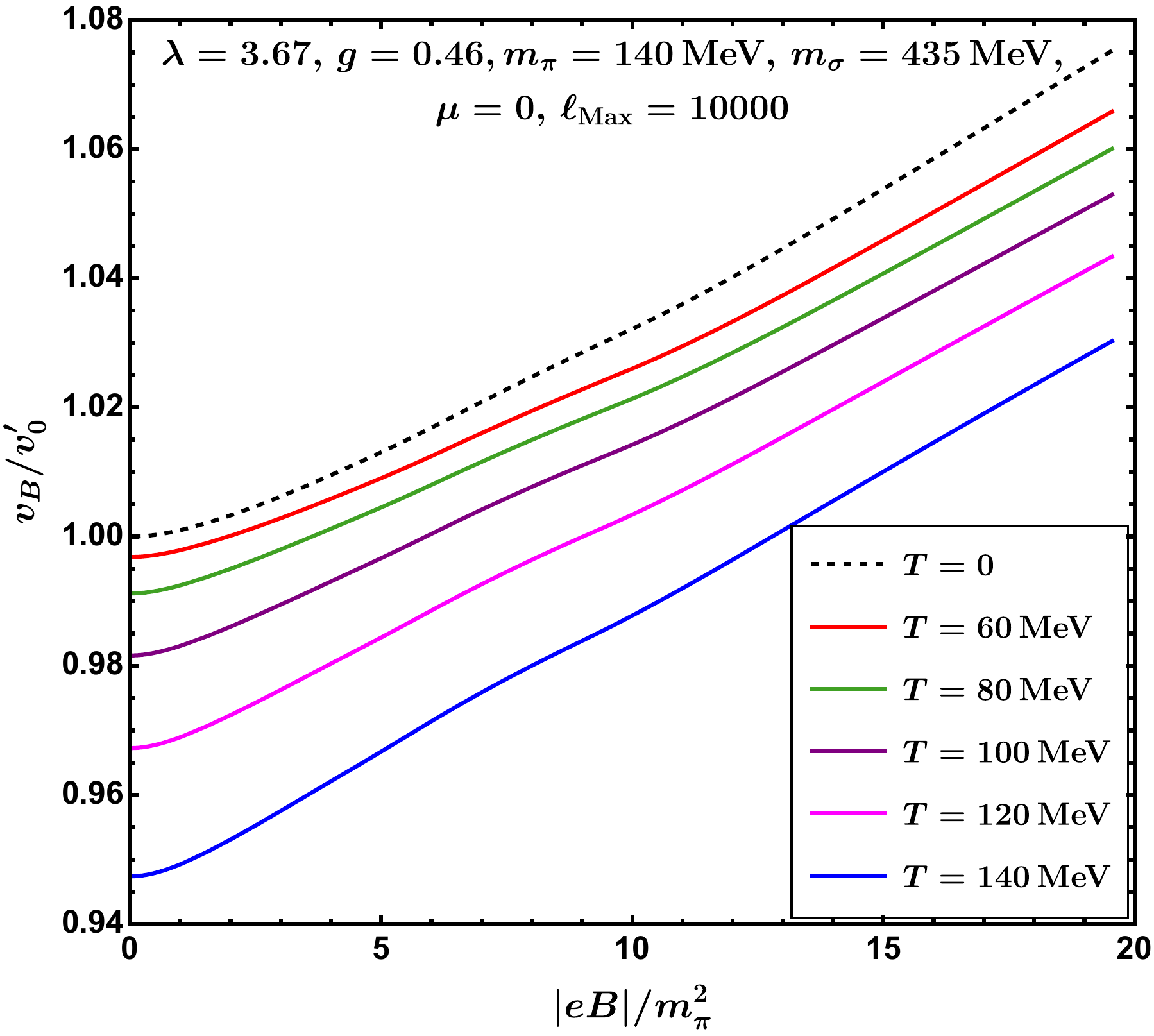}
		\includegraphics[scale=0.4]{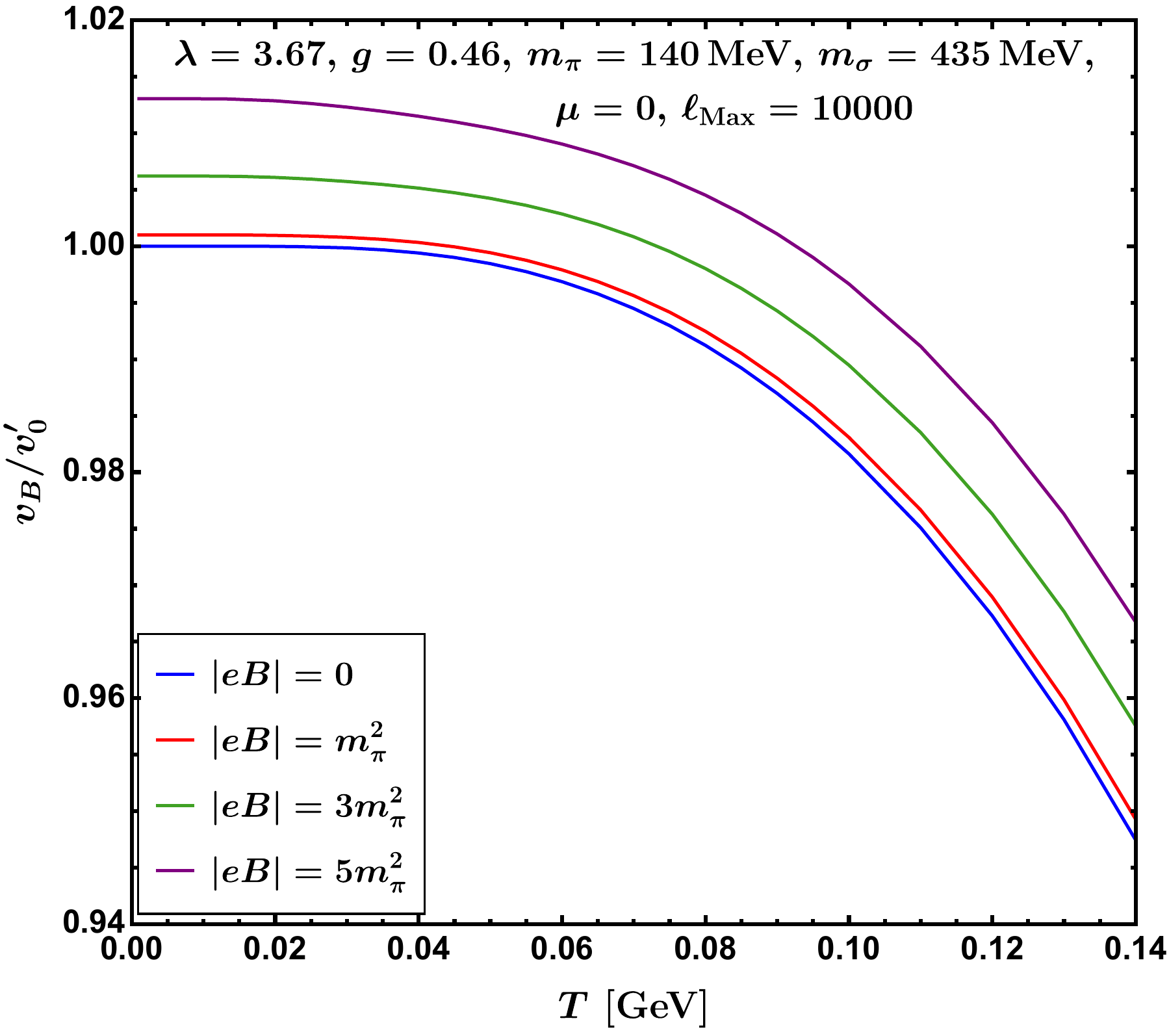}
		\caption{[color online] Figure shows the behavior of the minimum of the effective potential $v_B$ with $|eB|$ and $T$ shown on the left and right panels, respectively.}
	\end{figure}
	\section{Conclusion} \label{sec:conclusion}
	In conclusion, we have computed neutral pion mass in the presence of arbitrary background magnetic field at a non-zero temperature in the framework of LSMq model. We have examined the behavior of pion mass with the magnetic field, keeping the temperature at fixed values, and that with temperature, keeping the field fixed. The coupling constants of the theory are the meson-meson coupling $\lambda$ and quark-meson coupling $g$. At the same time, other parameters are vacuum pion mass $m_{\pi}$, vacuum sigma mass $m_{\sigma}$, and constituent quark mass $M_f$. We have incorporated the effect of meson self-coupling through $\lambda_{\text{eff}}$ and the quantum correction of effective potential through $v_{B}(T)$ in the $\pi^0$ mass. In calculating the mass, we have shown the magnetic field and temperature dependence of the one-loop effective potential $V_{\text{eff}}$ as a function of $v$. The general case of arbitrary strength of the magnetic field is considered by using the general expression of charged pion and quark propagator without invoking strong and/or weak field approximation. Also, a framework for extending the calculation in the finite density domain is incorporated by considering the constituent quark chemical potential $\mu$. We report the decrement of pion mass with the strength of the background magnetic field on which some LQCD and effective model studies agree. The increasing behavior of temperature agrees qualitatively with ChPt study of ref.~\cite{Andersen:2012zc}. To our knowledge, there is no LQCD simulation in the literature investigating the pion mass with the strength of the background magnetic field at non-zero temperature. So, our investigation of pion mass, upon the availability of lattice data at a non-zero temperature in the near future, can shed light on the predictability of LSMq framework.  
	\section{Acknowledgements}
	A.D would like to thank Arghya Mukherjee and Md Sabir Ali for useful discussions. AD is supported by school of Physical Sciences, NISER. NH is supported in part by the SERB-MATRICS under Grant No. MTR/2021/000939.  
	\appendix
	\section{Vertex correction} \label{sec:vertex}
\begin{figure}
		\centering
		\includegraphics[scale=0.4]{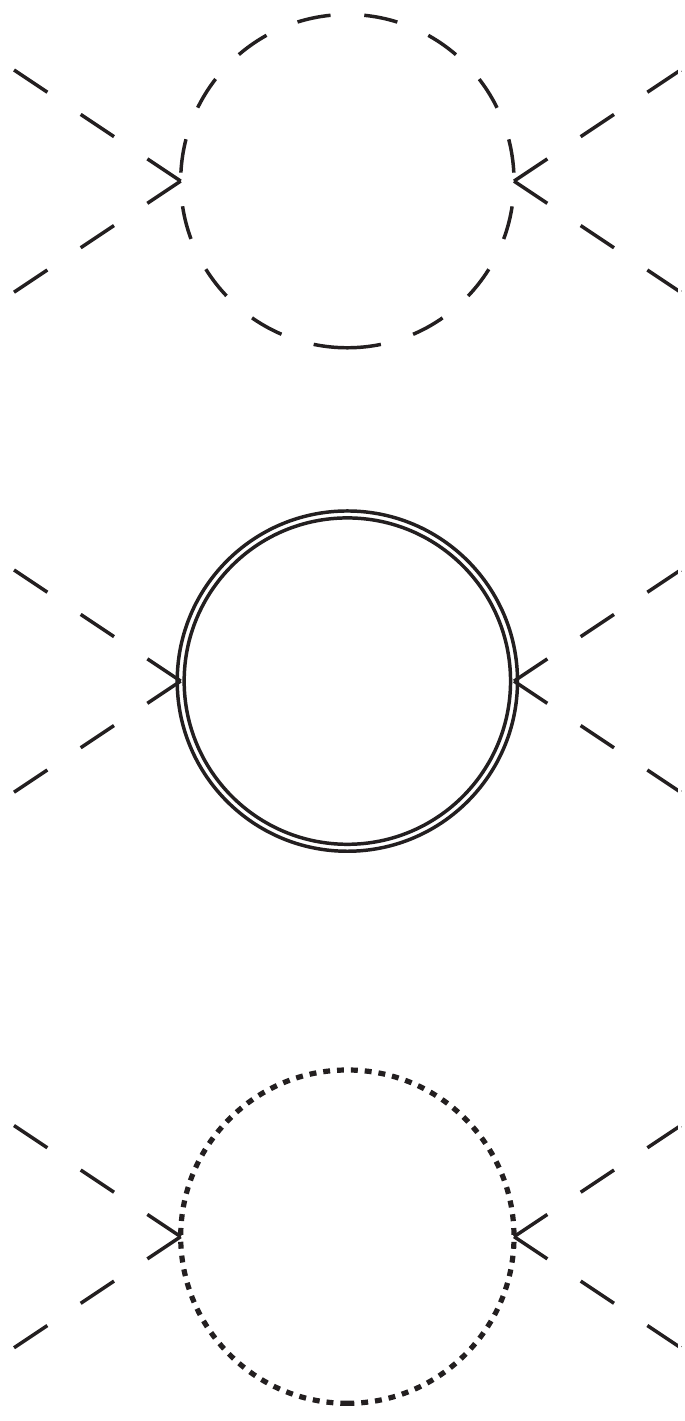}
		\caption{Feynman diagram for one-loop contribution to the self-coupling $\lambda$. The dashed line denotes $\pi^{0}$, double line denotes $\pi^{\pm}$, dotted line denotes $\sigma$-meson.}
		\label{fig:pi_vertex}
	\end{figure}

	The Feynman diagrams that contribute to the vertex correction of $\pi^0$ is depicted in Fig.~\ref{fig:pi_vertex}
	The expression for $\Delta \lambda$ is given by~\cite{Ayala:2018zat}
	\begin{align}
		\Delta \lambda = \frac{24\lambda^2}{16}\Big[9I(P,T,m_{\sigma})+I(P,T,m_{\pi})+4J^{\sB}(P,T,m_{\pi})\Big]\Bigg|_{\bm{p}\rightarrow \bm{0}}. \label{eq:delta_lambda}
	\end{align}
	As usual, $\Delta \lambda$ contains the magnetic vacuum part and thermo-magnetic part. Here we defined
	\begin{align}
		I(P,T,m_i) = T\sum_{k_0}\int\dfrac{d^3k}{(2\pi)^3}D_{i}(K)D_{i}(P-K), \nn
		J^{\sB}(P,T,m_j) = T\sum_{k_0}\int\dfrac{d^3k}{(2\pi)^3}D^{\sB}_{j}(K)D^{\sB}_{j}(P-K), \label{eq:I&J}
	\end{align} 
	with and $i=\pi^0,\sigma$ and $j=\pi^{+}, \pi^{-}$.
	\subsection{$\Delta\lambda_{B}$}
	There will be no magnetic vacuum part for $I$ since its expression contains only neutral pion and sigma loop. Only the $J$ contribution from the magnetic field will be there due to the involvement of the charged pion propagator inside the loop. Thus for the vacuum as well as magnetic field contributions, we write 
	\begin{align}
		J^{\sB}(P,m_j) &= -i\int\dfrac{d^4K}{(2\pi)^4}D^{\sB}_{j}(K)D^{\sB}_{j}(P-K)\nn
		&= -4i\sum_{\ell,n=0}^{\infty}(-1)^{\ell+n}\int\frac{d^2k_{\sperp}}{(2\pi)^2}\exp\left(-\frac{k_{\sperp}^2+q_{\sperp}^2}{|eB|}\right)L_{\ell}\left(\frac{2k_{\sperp}^2}{|eB|}\right)L_{n}\left(\frac{2q_{\sperp}^2}{|eB|}\right)\nn
		&\hspace{2cm}\times\int\frac{d^2k_{\shp}}{(2\pi)^2}\frac{1}{k_{\shp}^2-(2\ell+1)|eB|-m_{j}^2}\frac{1}{q_{\shp}^2-(2n+1)|eB|-m_j^2}
	\end{align}
	For $\bm{p}\rightarrow\bm{0}$, after performing the following perpendicular momentum integration and the sum over landau level $n$, the above Eq. gets simplified to 
	\begin{align}
		J^{\sB}(p_0,\bm{p}=0,m_j) &= -i\frac{|eB|}{2\pi}\sum_{\ell=0}^{\infty}\int\frac{d^2k_{\shp}}{(2\pi)^2}\frac{1}{k_{\shp}^2-(2\ell+1)|eB|-m_{j}^2}\frac{1}{q_0^2-k_z^2-(2l+1)|eB|-m_j^2}\nn
		&=-i\frac{|eB|}{2\pi}\sum_{\ell=0}^{\infty}\int\frac{d^2k_{\shp}}{(2\pi)^2}\int\limits_{0}^{1}dx\frac{1}{\left\{(k_0-x p_0)^2-k_z^2-\left[(2\ell+1)|eB|+m_{j}^2-x(1-x)p_{\shp}^2\right]\right\}^2}\nn
		&=-i\frac{|eB|}{2\pi}\sum_{\ell=0}^{\infty}\int\limits_{0}^{1}dx\int\frac{d^2k_{\shp}}{(2\pi)^2}\frac{1}{\left\{k_{\shp}^2-\left[(2\ell+1)|eB|+m_{j}^2-x(1-x)p_0^2\right]\right\}^2}
	\end{align}
	Now we perform the usual dimensional regularisation routine to get
	\begin{align}
		J^{\sB}(p_0,m_j) &=-i\frac{|eB|}{2\pi}\Lambda^{2\epsilon}\sum_{\ell=0}^{\infty}\int\limits_{0}^{1}dx\int\frac{d^{2-\epsilon}k_{\shp}}{(2\pi)^{2-\epsilon}}\frac{1}{\left\{k_{\shp}^2-\left[(2\ell+1)|eB|+m_{j}^2-x(1-x)p_{0}^2\right]\right\}^2}
	\end{align} 
	The momentum integration is performed as
	\begin{align}
		J^{\sB}(p_0,m_j) &=\frac{|eB|}{2\pi}\Lambda^{2\epsilon}\sum_{\ell=0}^{\infty}\int\limits_{0}^{1}dx \frac{1}{(4\pi)^{1-\epsilon}}\frac{\Gamma\left(1+\epsilon\right)}{\left[(2\ell+1)|eB|+m_j^2-x(1-x)p_{0}^2\right]^{1+\epsilon}}\nn
		&=\frac{|eB|}{8\pi^2}\left(\frac{1}{4\pi\Lambda^2}\right)^{-\epsilon}\Gamma\left(1+\epsilon\right)\int\limits_{0}^{1}dx\sum_{\ell=0}^{\infty}\frac{1}{\left[(2\ell+1)|eB|+m_j^2-x(1-x)p_{0}^2\right]^{1+\epsilon}} \nn
		&=\frac{1}{16\pi^2}\left(\frac{2|eB|}{4\pi\Lambda^2}\right)^{-\epsilon}\Gamma\left(1+\epsilon\right)\int\limits_{0}^{1}dx\,\,\zeta\left(1+\epsilon,\frac{1}{2}+\frac{m_j^2-x(1-x)p_{0}^2}{2|eB|}\right)
	\end{align}
	where we summed over LLs as
	\begin{align}
		\sum_{\ell=0}^{\infty}\frac{1}{\left[(2\ell+1)|eB|+m_j^2-x(1-x)p_0^2\right]^{1+\epsilon}} = \frac{1}{2|eB|}(2|eB|)^{-\epsilon}\zeta\left(1+\epsilon,\frac{1}{2}+\frac{m_j^2-x(1-x)p_0^2}{2|eB|}\right).
	\end{align}
	Finally, expanding $J^{\sB}$ in equation around $\epsilon=0$, we get
	\begin{align}
		J^{\sB}(p_0,\bm{p}=0,m_j) &= \frac{1}{16\pi^2}\left[\frac{1}{\epsilon}-\gamma_E+\log\left(4\pi\Lambda^2\right)-\log(2|eB|)-\int\limits_{0}^{1}dx\,\,\psi\left(\frac{1}{2}+\frac{m_j^2-x(1-x)p_{0}^2}{2|eB|}\right)\right].
	\end{align}
	\subsection{$\Delta\lambda_{\text{Vac}}$}
	Now, we compute the dimensionally regularised vacuum part by going to $d=4-2\epsilon$ as 
	\begin{align}I^{\text{Vac}}(p_0,\bm{p}=0,m_i) &= \frac{\lambda^2}{16\pi^2}\left[\frac{1}{\epsilon}-\gamma_E+\log(4\pi\Lambda^2)-\int\limits_{0}^{1}dx\log\left(m_i^2-x (1-x) p_0^2\right)\right], \nn
		J^{\text{Vac}}(p_0,\bm{p}=0,m_j) &= \frac{\lambda^2}{16\pi^2}\left[\frac{1}{\epsilon}-\gamma_E+\log(4\pi\Lambda^2)-\int\limits_{0}^{1}dx\log\left(m_j^2-x (1-x) p_0^2\right)\right] 
	\end{align} and
	\begin{align}
		J^{\sB}(p_0,\bm{p}=0,m_j)&-J^{\text{Vac}}(p_0,\bm{p}=0,m_j) \nn
		&= \frac{1}{16\pi^2}\int\limits_{0}^{1}dx\,\,\left[\log\left(\frac{m_j^2-x (1-x)p_0^2}{2|eB|}\right)-\psi\left(\frac{1}{2}+\frac{m_j^2-x (1-x)p_0^2}{2|eB|}\right)\right].
	\end{align}
	As long as $m_{\pi}^2-x (1-x) p_0^2 \geq 0$\footnote{To maintain this condition, we must choose $p^2_0<4m_{\pi}^2$ for the $J^{\sB}-J^{\text{Vac}}$ to be pure real}, the PolyGamma function in the above line can be expanded in the limit of $|eB|\rightarrow 0$ as
	\begin{align}
		\psi\left(\frac{1}{2}+\frac{m_j^2-x (1-x)p_0^2}{2|eB|}\right) = \log\left(\frac{m_j^2-x (1-x)p_0^2}{2|eB|}\right)+\frac{1}{24}\left(\frac{2|eB|}{m_j^2-x (1-x)p_0^2}\right)^2+\mathcal{O}\left(|eB|^4\right).
	\end{align}
	\subsection{$\Delta\lambda_{\text{ThM}}$}
	To extract the thermo-magnetic contribution, we start from the expression of $I$ and $J$ given in Eq.~\eqref{eq:I&J}. For $I$, we perform the frequency summation and drop the term that does not contain distribution function, and obtain the thermal part as 
	\begin{align}
		I^{\text{Th}}(p_0,\bm{p}=0,T,m_i) =\int\frac{d^3k}{(2\pi)^3}\frac{1}{E_{i,k}}\frac{1}{p_0^2-4E_{i,k}^2}2n\left(E_{i,k}\right),
	\end{align}
	where $E_{i,k}=\sqrt{k^2+m_i^2}$.\\
	
	For the thermo-magnetic part of vertex correction, we need to evaluate the following expression 
	\begin{align}
		J^{\sB}(p_0,\bm{p}=0,m_j) &= \frac{|eB|}{2\pi}T\sum_{k_0}\int\frac{dk_{z}}{2\pi}\sum_{\ell=0}^{\infty}\frac{1}{k_{\shp}^2-(2\ell+1)|eB|-m_{j}^2}\frac{1}{q_{\shp}^2-(2\ell+1)|eB|-m_j^2}.
	\end{align}
	The bosonic frequency sum is evaluated in Appendix~\ref{sec:freq_sums_boson}. In our case, we have the chemical potential of boson $\mu_b = 0$ and $\bm{p}=0$, giving
	\begin{align}
		T\sum_{k_0}\frac{1}{k^2_0-E^2_{k,\ell}}&\frac{1}{(p_0-k_0)^2-E^2_{j,k,\ell}} = -\frac{1+2n(E_{j,k,\ell})}{E_{j,k,\ell}\(p_0^2-4E^2_{j,k,\ell}\)}.
	\end{align}
	Finally, the thermo-magnetic part is written as
	\begin{align}
		J^{\text{ThM}}(p_0,\bm{p}=0,T,m_j) =-\frac{|eB|}{2\pi}\sum_{\ell = 0}^{\infty}\int\limits_{-\infty}^{\infty}\frac{dk_z}{2\pi}\frac{1}{E_{j,\ell,k}}\frac{1}{p_0^2-4E_{j,\ell,k}^2}2n\left(E_{j,\ell,k}\right),
	\end{align}
	where $E_{j,\ell,k}=\sqrt{k_z^2+(2\ell+1)|eB|+m_j^2}$. 
	\section{Frequency Sums}\label{sec:freq_sums}
	\subsection{Fermionic sums}\label{sec:freq_sums_fermionic}
	The sum of the form 
	\begin{align}
		&I_{s}=T\sum_{l=-\infty}^{\infty}\widetilde{\Delta}\big(k_0,s\Omega\big), \label{eq:I_s}\\
		&I_{s_1,s_2}=T\sum_{l=-\infty}^{\infty}\widetilde{\Delta}\big(k_0,s_1\Omega_1\big)\widetilde{\Delta}\big(q_0,s_2\Omega_2\big), \label{eq:I_s1_s2}
	\end{align}
	where
	\begin{align}
		\widetilde{\Delta}(k_0,s_1\Omega_1)\equiv \frac{s_1}{2\Omega_1}\frac{1}{k_0-s_1\Omega_1}, \\
		\widetilde{\Delta}(p_0-k_0, s_2\Omega_2) \equiv \frac{s_1}{2\Omega_1}\frac{1}{k_0-s_1\Omega_1},
	\end{align}
	with
	\begin{align}
		k_0=\mu + i (2l+1)\pi T,\,\,\, p_0 = i 2\pi m T,
	\end{align}
	where $l,m=0,\pm 1, \pm 2, \cdots$.
	Now, we substitute 
	\begin{align}
		&\frac{1}{i(2l+1)\pi T+\mu-s_1\Omega_1} = \widetilde{n}^{+}(s_1\Omega_1)\int_{0}^{1/T}d\tau_1\, \exp\left[-\tau_1\left\{i(2l+1)\pi T+\mu-s_1\Omega_1\right\}\right] \\
		&\frac{1}{i2\pi m T-i (2l+1)\pi T-\mu - s_2\Omega_2} = \widetilde{n}^{-}(s_2\Omega_2)\int_{0}^{1/T}d\tau_2\, \exp\left[-\tau_2\left\{i2\pi m T-i (2l+1)\pi T-\mu - s_2\Omega_2\right\}\right]
	\end{align} 
	in Eq.~\eqref{eq:I_s1_s2} and simplify the terms in the exponential to get
	\begin{align}
		I_{s_1,s_2} = \widetilde{n}_1^{+}\widetilde{n}_2^{-}\int_{0}^{1/T}d\tau_1 d\tau_2 e^{-\left(i2m\pi T \tau_2-s_1\omega_1\tau_1-s_2\Omega_2\tau_2\right)-\mu (\tau_1-\tau_2) }\,\,T\!\!\sum_{l=-\infty}^{\infty}\exp\left[-(\tau_1-\tau_2)i(2l+1)\pi T\right]. 
	\end{align}
	Here $\widetilde{n}_i^{\pm} \equiv \widetilde{n}^{\pm}(s_i\Omega_i) = \left( e^{\beta (s_iE_i\mp\mu)}+1 \right)^{-1}$, where $\beta = T^{-1}$. Now, we use the identity
	\begin{align}
		T\!\!\sum_{l=-\infty}^{\infty}\exp\left[-(\tau_1-\tau_2)i(2l+1)\pi T\right] = \delta\left(\tau_1-\tau_2\right)
	\end{align}
	to get
	\begin{align}
		I_{s_1,s_2} = \widetilde{n}_1^{+}\widetilde{n}_2^{-}\int_{0}^{1/T}d\tau \exp\left[-\tau \left(i2\pi m T-s_1\Omega_1-s_2\Omega_2\right)\right].
	\end{align}
	After performing the $\tau$ integral and simplifying by using $e^{i2m\pi}=1$, we analytically continue back to the Minkowski $p_0$ by the prescription $i2\pi m T\rightarrow p_0+i\varepsilon$. Then we make use of the identity $\widetilde{n}^{\pm}(E)e^{\beta (E\mp \mu)}=1-\widetilde{n}^{\pm}(E)$ and do a little algebra to arrive
	\begin{align} 
		I_{s_1,s_2} = -\frac{s_1s_2}{4\Omega_1\Omega_2}\frac{1-\widetilde{n}^{+}(s_1\Omega_1)-\widetilde{n}^{-}(s_2\Omega_2)}{p_0-s_1\Omega_1-s_2\Omega_2}.
	\end{align}
	Finally, using the identity $1-\widetilde{n}^{\pm}(-x)-\widetilde{n}^{\mp}(x)=0$, we get our desired frequency sum as 
	\begin{align}
		\sum_{s_1,s_2} I_{s_1,s_2} & = -\frac{1}{4\Omega_1\Omega_2}\left[\frac{1-\widetilde{n}^{+}(\Omega_1)-\widetilde{n}^{-}(\Omega_2)}{p_0-\Omega_1-\Omega_2}-\frac{1-\widetilde{n}^{-}(\Omega_1)-\widetilde{n}^{+}(\omega_2)}{p_0+\Omega_1+\Omega_2}+\frac{\widetilde{n}^{+}(\Omega_1)-\widetilde{n}^{+}(\Omega_2)}{p_0-\Omega_1+\Omega_2}-\frac{\widetilde{n}^{-}(\Omega_1)-\widetilde{n}^{-}(\Omega_2)}{p_0+\Omega_1-\Omega_2}\right].
	\end{align}
	Now, the following method can perform the other fermionic frequency sum as 
	\begin{align}
		& T\sum_{k_0}\frac{1}{k_0^2-\Omega^2} = T\sum_{k_0}\sum_{s=\pm 1}\frac{s}{2\Omega}\frac{1}{k_0-s\,\Omega} = \sum_{s=\pm 1}\frac{s}{2\Omega}\,T\!\!\sum_{l=-\infty}^{\infty}\frac{1}{i(2l+1)\pi T+\mu-s\,\Omega}\nn
		& = \sum_{s=\pm 1}\frac{s}{2\Omega}\widetilde{n}^{+}(s\,\Omega)\int_{0}^{1/T}d\tau e^{-\tau (\mu-s\,\Omega)}\,T\!\!\sum_{l=-\infty}^{\infty}e^{i(2l+1)\pi T \tau} = \sum_{s=\pm 1}\frac{s}{2\Omega}\widetilde{n}^{+}(s\,\Omega)\int_{0}^{1/T}d\tau e^{-\tau (\mu-s\,\Omega)}\delta (\tau) \nn
		&= \sum_{s=\pm 1}\frac{s}{2\Omega}\widetilde{n}^{+}(s\,\Omega).
	\end{align}
	Thus,
	\begin{align}
		T\sum_{k_0}\frac{1}{k_0^2-\Omega^2} = -\frac{1-\widetilde{n}^{+}(\Omega)-\widetilde{n}^{-}(\Omega)}{2\,\Omega}.
	\end{align}	 
	\subsection{Bosonic sums} \label{sec:freq_sums_boson}
	The frequency sum we need to evaluate is 
	\begin{align}
		\mathcal{F}_{B,B}^{(0,0)} = T\,\sum_{k_0}\frac{1}{k^2_0-E_1^2}\frac{1}{(p_0-k_0)^2-E^2_2}, \label{eq:fbb00}
	\end{align}
	where 
	\begin{align}
		k_0 = \mu+i2l\pi T.
	\end{align}
	The summand in Eq.~\eqref{eq:fbb00} can be conveniently written as
	\begin{align}
		\mathfrak{I}_{s_1,s_2} = T\sum_{\sss{l= -\infty}}^{\infty} \Delta\left(k_0,s_1E_1\right)\Delta\left(p_0-k_0,s_2E_2\right) \label{eq:Is1s2},
	\end{align} 
	where 
	\begin{align}
		&\Delta\(k_0,s_1E_1\) = \frac{s_1}{2E_1}\frac{1}{k_0-s_1E_1}, \nn
		&\Delta\(p_0-k_0,s_2E_2\) =  \frac{s_2}{2E_2}\frac{1}{p_0-k_0-s_2E_2}. \label{eq:delta_def}
	\end{align}
	Thus, Eq.~\eqref{eq:fbb00} can be written as
	\begin{align}
		\mathcal{F}_{B,B}^{(0,0)} = \sum_{\sss{s_1,s_2\atop =\pm 1}} \mathfrak{I}_{s_1,s_2}.
	\end{align}
	It is easy to see that Eq.~\eqref{eq:delta_def} can be written in integral representation as
	\begin{align}
		&\Delta\(k_0,s_1E_1\)= -\frac{s_1}{2E_1}n^{+}(s_1E_1)\int\limits_{0}^{1/T}d\tau_1 e^{-\tau_1 \(k_0-s_1E_1\)}, \nn
		&\Delta\(p_0-k_0,s_2E_2\)= -\frac{s_2}{2E_2}n^{-}(s_2E_2)\int\limits_{0}^{1/T}d\tau_2 e^{-\tau_2 \(p_0-k_0-s_2E_2\)}.
	\end{align}
	Thus,
	\begin{align}
		\mathfrak{I}_{s1,s2} &= \frac{s_1s_2n^{+}(s_1E_1)n^{-}(s_2E_2)}{4E_1E_2} \int_{0}^{1/T} d\tau_1 d\tau_2\,\,e^{\tau_1\left(s_1E_1-\mu\right)}e^{\tau_2\left(s_2E_2+\mu\right)-\tau_2p_0}\times T\sum_{\sss{l=-\infty}}^{\infty}\exp\left[-k_0\left(\tau_1-\tau_2\right)\right].
	\end{align}
	Using the identity
	\begin{align}
		T\sum_{\sss{l=-\infty}}^{\infty}\exp\left[-k_0\(\tau_1-\tau_2\)\right] = \delta\( \tau_1 - \tau_2 \),
	\end{align}
	and integrating over the delta function
	\begin{align}
		\mathfrak{I}_{s1,s2} &= \frac{s_1s_2n^{+}(s_1E_1)n^{-}(s_2E_2)}{4E_1E_2} \int_{0}^{1/T} d\tau \,\,e^{\tau\left(s_1E_1-\mu\right)}e^{\tau\left(s_2E_2+\mu\right)}e^{-\tau p_0}.
	\end{align}
	Performing the $\tau$ integral, we get
	\begin{align}
		&\mathfrak{I}_{s1,s2} =\frac{s_1s_2}{4E_1E_2}\,n^{+}(s_1E_1)n^{-}(s_2E_2) \dfrac{e^{\beta\left(s_1E_1-\mu\right)}e^{\beta\left(s_2E_2+\mu\right)}e^{-\beta p_0}-1}{s_1E_1+s_2E_2-p_0},
	\end{align} 
	Since $e^{-\beta p_0}=1$, we get after some algebra 
	\begin{align}
		\mathfrak{I}_{s1,s2} = -\frac{s_1s_2}{4E_1E_2}\dfrac{1+n^{+}(s_1E_1)+n^{-}(s_2E_2)}{p_0-s_1E_1-s_2E_2}.
	\end{align}
	Using $n^{\pm}(E)=-\big[1+n^{\mp}\left(E\right)\big]$, we get
	\begin{align}
		&\mathcal{F}_{B,B}^{(0,0)} = \nn &-\frac{1}{4E_1E_2}\left(\frac{1+n^+(E_1)+n^-(E_2)}{p_0-E_1-E_2}-\frac{1+n^-(E_1)+n^+(E_2)}{p_0+E_1+E_2}-\frac{n^+(E_1)-n^+(E_2)}{p_0-E_1+E_2} +\frac{n^-(E_1)-n^-(E_2)}{p_0+E_1-E_2}\right).
	\end{align}
	\section{Effective Potential at Non-zero Temperature} \label{app:effective_potential}
	The effective potential is a central quantity for theories with a spontaneous breakdown of continuous symmetry. In this case, the classical value of potential is altered due to perturbative loop correction after spontaneous symmetry breaking. As a result of this, the minimum $v_0^{\prime} = \sqrt{(a^2+m_{\pi}^2)/\lambda}$ of tree level potential $V^{\prime}_{\text{tree}}(v)$ receives quantum correction shifting it's value to $v = v_{B}$. In the lowest order in perturbation theory, the effective potential is just classical potential.    
	In this section, we compute the contribution of temperature and magnetic field to the effective potential. First, the effective potential has contributions from tree level, bosonic (appearing due to the quantum fluctuations of $\pi$ and $\sigma$ meson), and fermionic part (for which quantum fluctuation of quarks are responsible). The higher-order corrections to the potential are divergent, and the incorporation of counterterm contribution $V_{\text{ct}}$ is needed to remove the infinities systematically. Thus up to $\mathcal{O}(\hbar)$, it reads
	\begin{equation}
		V_{\text{eff}} = V^{\prime}_{\text{tree}} + \sum_{b = \pi^{\pm}, \pi^0, \sigma}V_{b}^{(1)} + \sum_{f=u,d}V_{f}^{(1)} + V_{\text{ct}} + \sum_{b = \pi^{\pm}, \pi^0, \sigma}V_{b,\text{Ring}}^{(1)},
	\end{equation}
	where 
	\begin{align}
		V^{\prime}_{\text{tree}} =&-\frac{1}{2}(a^2+m_{\pi}^2)v^2+\frac{1}{4}\lambda v^4, \\
		V_b^{(1,B)}=-&\frac{1}{2}T\sum_{n=-\infty}^{\infty}\int\frac{d^3k}{(2\pi)^3}\log\left[D_{\sB}(k_0 = i\omega_n,\bm{k},m^2_b)^{-1}\right], \label{eq:Vb_org_def}\\
		V_f^{(1,B)} = &\frac{i}{2}\textsf{Tr}\log\left[(i\slashed{D})^2-M_{f}^2\right], \label{eq:Vf_org_def}\\
		V_{\text{ct}}=&\frac{1}{2}\delta m \,\,v^2+\frac{1}{4}\delta \lambda \,\,v^4
	\end{align}
	and $V_{b,\text{Ring}}^{(1)}$ is the Ring contribution from mesons which is discussed in detail in Sec.\ref{subsec:Ring}. We shall not write the $v$ dependence, which is there in the expression of effective potential via $m_b^2$ and $M_{f}$.  \\
	After few steps of simple algebra clearly depicted in ref~\cite{Ayala:2021nhx}, Eq.~\eqref{eq:Vb_org_def} and Eq.~\eqref{eq:Vf_org_def} takes the following form
	\begin{align}
		V_b^{(1,B)}=&\frac{1}{2}T\sum_{n=-\infty}^{\infty}\int_{0}^{\infty}dm_b^2\int\frac{d^3k}{(2\pi)^3}\int_{0}^{\infty}ds\frac{1}{\cosh(|eB|)}\exp\(-s\bigg[\omega^2_{n}+k_z^2+\frac{\tanh(|eB|s)}{|eB|s}k_{\sperp}^2+m_{b}^2\bigg]\),\label{eq:Vb_proper_time}\\
		V_f^{(1,B)}=-&\sum_{r=\pm 1}T\sum_{n=-\infty}^{\infty}\int_{0}^{\infty}dm_f^2\int\frac{d^3k}{(2\pi)^3}\int_{0}^{\infty}\frac{ds}{\cosh(|q_{f}B|)}\exp\(-s\bigg[\tilde{\omega}^2_{n}+k_z^2+\frac{\tanh(|q_{f}B|s)}{|q_{f}B|s}k_{\sperp}^2+M_{f}^2+r\,q_{f}B\bigg]\).\label{eq:Vf_proper_time}
	\end{align}
	Here, $\omega_n = 2\pi nT$ and $\widetilde{\omega}_n=(2n+1)\pi T-i\mu$ are bosonic and fermionic Matsubara frequencies, respectively.
	By integrating over the proper time $s$ in Eq.~\eqref{eq:Vb_proper_time} and \eqref{eq:Vf_proper_time}, the expressions of $V_b^{(1)}$, $V_f^{(1)}$ is converted to the Landau level representation.
	\begin{align}
		V_{b}^{(1,B)}&= \frac{T}{2} \sum_{n=-\infty}^{\infty} \int d m_{b}^{2} \int \frac{d^{3} k}{(2 \pi)^{3}} \sum_{\ell = 0}^{\infty} 2(-1)^{\ell}\frac{\exp\left(-\frac{k_{\sperp}^2}{|eB|}\right)L_{\ell}\left(\frac{2k_{\sperp}^2}{|eB|}\right)}{\omega_{n}^2+k_z^2+(2\ell+1)|eB|+m_b^2}, \label{eq:Vb_landau_level}\\
		V_{f}^{(1,B)} &= -\sum_{r=\pm 1} T \sum_{n=-\infty}^{\infty} \int d m_{f}^{2} \int \frac{d^{3} k}{(2 \pi)^{3}} \sum_{\ell= 0}^{\infty} 2(-1)^{\ell}\frac{\exp\left(-\frac{k_{\sperp}^2}{|eB|}\right)L_{\ell}\left(\frac{2k_{\sperp}^2}{|q_{f}B|}\right)}{\widetilde{\omega}_{n}^2+k_z^2+\left(2\ell+1+r\,\textsf{sgn}(q_{f}B)\right)|q_{f}B|+M_f^2}. \label{eq:Vf_landau_level}
	\end{align}
	After performing the sum over $r$, Eq.~\eqref{eq:Vf_landau_level} can be straightforwardly simplified by writing in terms of spin degeneracy factor as
	\begin{equation}
		V_{f}^{(1,B)} = - T \sum_{n=-\infty}^{\infty} \int d m_{f}^{2} \int \frac{d^{3} k}{(2 \pi)^{3}} \sum_{\ell = 0}^{\infty} (2-\delta_{l,0})2(-1)^{\ell} \frac{\exp\left(-\frac{k_{\sperp}^2}{|eB|}\right)L_{\ell}\left(\frac{2k_{\sperp}^2}{|q_{f}B|}\right)}{\widetilde{\omega}_{n}^2+k_z^2+2\ell|q_{f}B|+M_f^2}. \label{eq:Vf_landau_level_2nd}
	\end{equation} 
	\subsection{Computation of $V_b^{(1)}$}
	By performing the integration over $d^{2} k_{\perp},$  we get
	\begin{align}
		V_{b}^{(1,B)}&= T \sum_{n=-\infty}^{\infty} \int d m_{b}^{2}\int_{-\infty}^{\infty}\frac{dk_{z}}{2\pi}\frac{|eB|}{4\pi}\sum_{\ell = 0}^{\infty}\frac{1}{\omega_{n}^2+k_z^2+(2\ell+1)|eB|+m_b^2}, \nn
		&=\frac{|eB|}{4\pi} \int_{-\infty}^{\infty}\frac{dk_{z}}{2\pi}\sum_{\ell = 0}^{\infty}\,\,\,\, T\!\!\sum_{n=-\infty}^{\infty}\log\left[\omega_{n}^2+k_z^2+(2\ell+1)|eB|+m_b^2\right]. \label{eq:V_b^1_sum}
	\end{align}
	Now the frequency sum is performed following Ref.~\cite{Bellac:2011kqa} as
	\begin{align}
		T\!\!\sum_{n=-\infty}^{\infty}\log\left(\omega_{n}^2+E_{k,\ell,b}^2\right) =E_{k,\ell,b}+2 T \log\left[1-\exp\left(-\frac{E_{k,\ell,b}}{T}\right)\right] . 
	\end{align}
	Substituting the above expression of the sum integration in Eq.~\eqref{eq:V_b^1_sum}, we get,
	\begin{align}
		V_{b}^{(1,B)}&= \frac{|eB|}{4\pi} \sum_{\ell = 0}^{\infty}\int_{-\infty}^{\infty}\frac{dk_{z}}{2\pi}\left\{E_{\ell,k}+2T \log\left[1-\exp\left(-\frac{E_{k,\ell,b}}{T}\right)\right]\right\}.
	\end{align}
	Now the first term containing $E_{k,\ell,b}$ is divergent, which we need to regulate. We use the following procedure to regulate the momentum integration by dimensional regularization 
	\begin{align}
		\int_{-\infty}^{\infty}\frac{dk_{z}}{2\pi} \rightarrow \tilde{\Lambda}^{2\epsilon}\int\frac{d^{1-2\epsilon}k}{(2\pi)^{1-2\epsilon}}.
	\end{align} 
	We use the following identity in Ref.~\cite{Laine:2016hma}
	\begin{align}
		\Phi(m,d,A) = \int\frac{d^dk}{(2\pi)^d}\frac{1}{(\bm{k}^2+m^2)^A} = \frac{1}{(4\pi)^{d/2}}\frac{1}{\Gamma(A)}\Gamma\left(A-\frac{d}{2}\right)\frac{1}{(m^2)^{A-\frac{d}{2}}}
	\end{align}
	to perform the integration
	\begin{align}
		V_{b,\epsilon}^{(1,B)} &\equiv \frac{|eB|}{4\pi}\tilde{\Lambda}^{2\epsilon}\sum_{\ell = 0}^{\infty}\int\frac{d^{1-2\epsilon}k_z}{(2\pi)^{1-2\epsilon}} E_{k,\ell,b}=\frac{|eB|}{4\pi}\tilde{\Lambda}^{2\epsilon}\sum_{\ell = 0}^{\infty}\Phi\left(\sqrt{(2\ell+1)|eB|+m_b^2},1-2\epsilon,-\frac{1}{2}\right)\nn 
		&=- \frac{|eB|}{16\pi^2}\frac{\Gamma(\epsilon-1)}{\left(4\pi \tilde{\Lambda}^{2}\right)^{-\epsilon}} \sum_{\ell = 0}^{\infty} \frac{1}{\left[(2\ell+1)|eB|+m_b^2\right]^{\epsilon-1}}.
	\end{align}
	The sum over LL is performed by using the representation of Hurwitz zeta function
	\begin{align}
		\sum_{\ell=0}^{\infty} \frac{1}{(\ell+a)^{\epsilon}} = \zeta (\epsilon, a) 
	\end{align} 
	as
	\begin{align}
		V_{b,\epsilon}^{(1,B)}   = -\frac{|eB|^2}{8\pi^2}\left(\frac{2|eB|}{4\pi \tilde{\Lambda}^{2}}\right)^{-\epsilon}\Gamma\left(\epsilon - 1\right)\zeta\left(\epsilon -1, \frac{1}{2}+\frac{m_b^2}{2|eB|}\right)
	\end{align}
	Here we used $\Gamma(-1/2)=-\sqrt{4\pi}$. This result matches exactly with~\cite{Ayala:2014mla}. Now, we expand $V_{b,\epsilon}^{(1,B)} $ around $\epsilon = 0$ and obtain
		\begin{align}
			V_{b,\epsilon}^{(1,B)}  = \frac{|eB|^2}{8\pi^2}\left\{\zeta\left(-1,\frac{1}{2}+\frac{m_b^2}{2|eB|}\right)\left(\frac{1}{\epsilon}-\gamma_E-\log\frac{2|eB|}{4\pi \tilde{\Lambda}^{2}}+1\right)+\zeta^{(1,0)}\left(-1,\frac{1}{2}+\frac{m_b^2}{2|eB|}\right)\right\}+\mathcal{O}(\epsilon) \label{eq:Vbepsilon_1B}
		\end{align}
	The finite temperature part is
		\begin{align}
			V^{(1,B)}_{b,\text{ThM}} =  \frac{|eB|}{2\pi^2}T\sum_{l=0}^{\infty}\int\limits^{\infty}_{0}dk_z\,\,\log\left(1-e^{-E_{k,\ell,b}/T}\right). \label{eq:VbThM_1B}
	\end{align}
	For the $\sigma$ and $\pi^0$ meson, we take the limit of $|eB|\rightarrow 0$ in Eq.~\eqref{eq:Vb_proper_time} and do the $s$ and $m_b^2$ integration to get 
	\begin{align}
		V^{(1,B=0)}_{b}&=\frac{T}{2}\sum_{n=-\infty}^{\infty}\int\frac{d^3k}{(2\pi)^3}\log\left(\omega_n^2+k^2+m_b^2\right)\nn
		&=\frac{1}{2}\int\frac{d^3k}{(2\pi)^3}\left\{\sqrt{k^2+m_b^2}+2 T\log\left[1-\exp\left(-\frac{\sqrt{k^2+m_b^2}}{T}\right)\right]\right\}.
	\end{align}
	For the vacuum part, we use the dimensional regularisation method by modifying the measure of 3-momentum integration
	\begin{align}
		\int\frac{d^3k}{(2\pi)^3}\rightarrow \Lambda^{2\epsilon}\int\frac{d^{3-2\epsilon}k}{(2\pi)^{3-2\epsilon}}.
	\end{align} 
	As a result, we get,
		\begin{align}
			V^{(1,B=0)}_{b,\epsilon} &= \frac{1}{2}\Lambda^{2\epsilon}\int\frac{d^{3-2\epsilon}k}{(2\pi)^{3-2\epsilon}}\sqrt{k^2+m_b^2} = -\frac{m_b^2}{32\pi^2}\left(\frac{m_b^2}{4\pi\Lambda^2}\right)^{-\epsilon}\Gamma(\epsilon-2)\nn
			&=-\frac{m_b^4}{64\pi^2}\left[\frac{1}{\epsilon}-\gamma_E+\log(4\pi)+\frac{3}{2}-\log\left(\frac{m_b^2}{\Lambda^2}\right)\right]+ \mathcal{O}(\epsilon). \label{eq:Vbepsilon_1B=0}
	\end{align}
	The thermal part is given by
		\begin{align}
			V^{(1,B=0)}_{b,\text{Th}} &= \frac{T}{2\pi^2}\int\limits_{0}^{\infty}dk\,k^2\,\log\left[1-\exp\left(-\frac{\sqrt{k^2+m_b^2}}{T}\right)\right]. \label{eq:VbTh_1B=0}
	\end{align}
	\subsection{Computation of $V_f^{(1)}$}
	In this case, we start from Eq.~\eqref{eq:Vf_landau_level_2nd}
	\begin{align}
		V_{f}^{(1)} &= -2\sum_{\ell=0}^{\infty}(2-\delta_{\ell,0})T\sum_{n=-\infty}^{\infty}\int\frac{d^3k}{(2\pi)}(-1)^l\exp\left(-\frac{k_{\sperp}^2}{|eB|}\right)L_{\ell}\left(\frac{2k_{\shp}^2}{|q_{f}B|}\right)\log\left[\widetilde{\omega}_n^2+k_z^2+2l|q_{f}B|+M_{f}^2\right]\nn
		&=-\frac{|q_{f}B|}{2\pi}\sum_{l=0}^{\infty}(2-\delta_{\ell,0})\int_{-\infty}^{\infty}\frac{dk_z}{2\pi}T\sum_{n=-\infty}^{\infty}\log\left[\widetilde{\omega}_n^2+k_z^2+2\ell|q_{f}B|+M_{f}^2\right].
	\end{align}
	
	The frequency sum is performed following the same method as for the bosonic part, and the result is quoted below
	\begin{align}
		T\sum_{n=-\infty}^{\infty} \log\left[\widetilde{\omega}_{n}^2+\Omega_{\ell,k,f}^2\right] = \Omega_{\ell,k,f}+T\log\left[1+\exp\left(-\frac{\Omega_{\ell,k,f}-\mu}{T}\right)\right]+T\log\left[1+\exp\left(-\frac{\Omega_{\ell,k,f}+\mu}{T}\right)\right]
	\end{align}
	Thus, we have
	\begin{align}
		V_{f}^{(1)} = -\frac{|q_{f}B|}{2\pi}\sum_{\ell=0}^{\infty}(2-\delta_{\ell,0})\int_{-\infty}^{\infty}\frac{dk_z}{2\pi} \bBigg@{2.5}\{\Omega_{\ell,k,f}+T\log\left[1+\exp\left(-\frac{\Omega_{\ell,k,f}-\mu}{T}\right)\right] \nn
		+\ T\log\left[1+\exp\left(-\frac{\Omega_{\ell,k,f}+\mu}{T}\right)\right] \bBigg@{2.5}\}.
	\end{align}
	The zero temperature part is written as
	\begin{align}
		V_{f,\epsilon}^{(1,B)} = -\frac{|q_fB|}{2\pi}\tilde{\Lambda}^{2\epsilon}\sum_{\ell=0}^{\infty} \left(2-\delta_{\ell,0}\right) \int\frac{d^{1-2\epsilon}k}{(2\pi)^{1-2\epsilon}}\,\,\Omega_{\ell,k,f}
	\end{align}
	The sum is performed, followed by the integration 
	\begin{align}
		V_{f,\epsilon}^{(1,B)} &=\frac{|q_{f}B|^2}{2\pi^2}\Gamma(\epsilon-1)\left[\left(\frac{2|q_{f}B|}{4\pi\Lambda^2}\right)^{-\epsilon}\zeta\left(\epsilon-1,\frac{M_{f}^2}{2|q_{f}B|}\right)-\frac{1}{2}\frac{M_{f}^2}{2|q_{f}B|}\left(\frac{M_{f}^2}{4\pi\Lambda^2}\right)^{-\epsilon}\right]\nn
		&=M_{f}^2\frac{|q_{f}B|}{8\pi^2}\left[\frac{1}{\epsilon}-\gamma_E-\log\left(\frac{M_{f}^2}{4\pi\Lambda^2}\right)+1\right]-\frac{|q_{f}B|^2}{2\pi^2}\left\{\zeta\left(-1,\frac{M_{f}^2}{2|q_{f}B|}\right)\left[\frac{1}{\epsilon}-\gamma_E-\log\left(\frac{2|q_{f}B|}{4\pi\Lambda^2}\right)+1\right]\right.\nn
		&\hspace{8cm}\left.+\zeta^{(1,0)}\left(-1,\frac{M_{f}^2}{2|q_{f}B|}\right)\right\}+\mathcal{O}(\epsilon).
	\end{align}
	After applying the $\overline{\text{MS}}$ scheme, by virtue of which we drop $\displaystyle \frac{1}{\epsilon}-\gamma_E+\log(4\pi)$ term, we get
		\begin{align}
			V_{f}^{(1,B)} &= \frac{|q_{f}B|}{2\pi^2}\bBigg@{3}\{\frac{M_{f}^2}{4}\left[1-\log\left(\frac{M_{f}^2}{\Lambda^2}\right)\right]\nn
			&\hspace{1cm}-|q_{f}B|\left[\zeta\left(-1,\frac{M_{f}^2}{2|q_{f}B|}\right)\left[1-\log\left(\frac{2|q_{f}B|}{\Lambda^2}\right)\right]+\zeta^{(1,0)}\left(-1,\frac{M_{f}^2}{2|q_{f}B|}\right)\right]\bBigg@{3}\}. \label{eq:Vfepsilon_1B}
	\end{align}
	The thermo-magnetic part is written as
		\begin{align}
			V_{f,\text{ThM}}^{(1,B)} =-T\frac{|q_{f}B|}{2\pi^2}\sum_{\ell=0}^{\infty}(2-\delta_{\ell,0})\int_{0}^{\infty}dk_z\left\{\log\left[1+\exp\left(-\frac{\Omega_{\ell,k,f}-\mu}{T}\right)\right]+\log\left[1+\exp\left(-\frac{\Omega_{\ell,k,f}+\mu}{T}\right)\right]\right\}. \label{eq:VfThM_1B}
		\end{align}
	The counter-terms are determined from vacuum stability condition~\cite{Carrington:1991hz}. It states that the tree level value of the position of minimum $v_{0}^{\prime}$ of effective potential and the mass of Sigma meson~\footnote{Note that the mass of sigma meson is equal to $\dfrac{d^2V_{\text{cl}}}{dv^2}\Big\vert_{v=v_{0}^{\prime}}$} does not change after quantum correction. Mathematically,
	\begin{align}
		\frac{1}{2v}\frac{dV_{\text{vac}}}{dv}\Big\vert_{v=v_0^{\prime}} &= 0,\nn
		\dfrac{d^2V_{\text{vac}}}{dv^2}\Big\vert_{v=v_{0}^{\prime}} &= 2a^2+3m_{\pi}^2
	\end{align}
	Applying the above conditions to the quantum corrected potential in the vacuum, we determine $\delta a^2$ and $\delta \lambda$ as 
	\begin{align}
		\delta a^2 &= \frac{m_{\pi}^2}{2}-\frac{1}{16\pi^2\lambda}\left\{6\lambda^2 (a^2+2m_{\pi}^2)-g^4(a^2+ m_{\pi}^2)+3 a^2 \lambda^2\left[\log\left(\frac{m_{\pi}^2}{a^2}\right)+\log\left(\frac{2a^2+3m_{\pi}^2}{a^2}\right)\right]\right\} \nn
		\delta \lambda &= \frac{\lambda}{2}\frac{m_{\pi}^2}{a^2+m_{\pi}^2}-\frac{1}{16\pi^2}\left\{3\lambda^2\left[\log\left(\frac{m_{\pi}^2}{a^2}\right)+3\log\left(\frac{2a^2+3m_{\pi}^2}{a^2}\right)\right]-8g^4\log\left(\frac{g^2}{a^2}\frac{a^2+m_{\pi}^2}{\lambda}\right)\right\}
	\end{align} 
	\subsection{Ring Contributions} \label{subsec:Ring}
	If we look at Eq.~\eqref{eq:Vbepsilon_1B}, Eq.~\eqref{eq:VbThM_1B}, Eq.~\eqref{eq:Vbepsilon_1B=0} and Eq.~\eqref{eq:VbTh_1B=0}, we notice that the argument of logarithm in Eq.~\eqref{eq:Vbepsilon_1B=0}, the argument of Hurwitch zeta function $\zeta$ in Eq.~\eqref{eq:Vbepsilon_1B}, Eq.~\eqref{eq:Vfepsilon_1B} can become negative due to negative $m_b^2$ for some values of $v$ in the range $0 < v < v_{0}^{\prime}$. This negative argument makes the potential imaginary which is not acceptable as it can lead to complex critical temperature $(T_{\text{chiral}})$ of chiral symmetry restoration~\cite{Dolan:1973qd}. Also, the meson energy $E_{\ell,k}$ and $E_{k}$ can become negative for a similar reason. \\
	
	Also, in the case of small boson mass, their thermal, as well as magnetic correction, becomes of the same order as their original masses. As a result, perturbation theory breaks down, and a resummation becomes necessary. It is taken into account by incorporating the so-called \emph{ring diagrams} as shown in Fig. By incorporating the resummation program through the inclusion of ring diagram, one takes the effect of plasma screening into account as well shields the effect of infra-red divergence. The ring diagram contribution is added via the following term~\cite{Ayala:2021nhx} 
	\begin{align}
		V^{(1,B)}_{b,\text{Ring}} = \frac{1}{2}T\!\!\sum_{n=-\infty}^{\infty}\int\frac{d^3k}{(2\pi)^3}\log\left[1-\Pi^{\sB}_b(k_0=i\omega_n,\bm{k})D_B(k_0=i\omega_n,\bm{k},m_b)\right]. \label{eq:Vb_Ring_org_def}
	\end{align}
	Now  adding Eq.~\eqref{eq:Vb_org_def} and Eq.~\eqref{eq:Vb_Ring_org_def}, we get,
	\begin{align}
		V_{b}^{(1,B)} + V_{b,\text{Ring}}^{(1,B)} = \frac{1}{2}T\!\!\!\sum_{n=-\infty}^{\infty}\int\frac{d^3k}{(2\pi)^3} \log\left[D_B(k_0=i\omega_n,\bm{k},m_b)^{-1}-\Pi^{\sB}_b(k_0=i\omega_n,\bm{k})\right]. \label{eq:Vb_(Ring+tree)_org_def}
	\end{align}
	Here, we rewrite the full expressions of meson self-energies as follows
	\begin{align}
		\Pi^{\sB}_{\pi^0}(k_0,\bm{k}) &= \frac{\lambda}{4}\left[8\mathcal{I}^{\sB}\left(m_{\pi^{\pm}}\right)+12\mathcal{I}\left(m_{\pi^0}\right)+4\mathcal{I}\left(m_{\sigma}\right)\right] + \sum_{f=u,d}\Pi_{ff}^{\sB}\left(k_0,\bm{k}\right), \\
		\Pi^{\sB}_{\pi^{\pm}}(k_0,\bm{k}) &= \frac{\lambda}{4}\left[16 \mathcal{I}^{\sB}\left(m_{\pi^{\pm}}\right)+4\mathcal{I}\left(m_{\pi^0}\right)+4\mathcal{I}\left(m_{\sigma}\right)\right] + 2 \Pi_{ud}^{\sB}\left(k_0,\bm{k}\right), \\
		\Pi^{\sB}_{\sigma}(k_0,\bm{k}) &= \frac{\lambda}{4}\left[8\mathcal{I}^{\sB}\left(m_{\pi^{\pm}}\right)+4\mathcal{I}\left(m_{\pi^{0}}\right)+12\mathcal{I}\left(m_{\sigma}\right)\right] + 2\Pi_{ud}^{\sB}\left(k_0,\bm{k}\right),
	\end{align}
	where the integrations $\mathcal{I}^{\sB}$ and $\mathcal{I}$ are defined as
	\begin{align}
		\mathcal{I}^{\sB}(m_i) &= T\sum_{k_0}\int\frac{d^3k}{(2\pi)^3}D^{\sB}(k_0,\bm{k}), \\
		\mathcal{I}(m_i) &= T\sum_{k_0}\int\frac{d^3k}{(2\pi)^3}D(k_0,\bm{k}).
	\end{align}
	The computation of the r.h.s of \eqref{eq:Vb_Ring_org_def} is analytically very challenging and numerically cumbersome. It is challenging to separate and regulate divergent contributions. Nevertheless, we can tackle it by invoking some educated approximations 
	\begin{itemize}
		\item First, we discard the $(k_0,\bm{k})$ dependency of $\Pi^{\sB}_b(k_0=i\omega_n,\bm{k})$ and consider the static limit, i.e, we take $\Pi^{\sB}_b(k_0=i\omega_n,\bm{k}) \simeq \Pi^{\sB}_b(k_0 = 0,\bm{k} \rightarrow 0) $.
		\item Next, we observe that computing the right-hand side Eq.~\eqref{eq:Vb_(Ring+tree)_org_def}  is the same as computing $V_b^{(1,B)}$ defined in Eq.~\eqref{eq:Vb_org_def} with $m_b^2$ being replaced by $m_b^2 + \Pi^{\sB}_b(k_0 = 0,\bm{k} \rightarrow 0) $ to an excellent approximation.
		
		\item We have taken $\Pi^{\sB}_{\pi^0} \approxeq \Pi^{\sB}_{\pi^{\pm}} \approxeq \Pi^{\sB}_{\sigma}$ since their order of magnitude is more or less the same.
	\end{itemize}
	
	Thus, after substituting $m_b^2 \rightarrow m_b^2 + \Pi$ in Eq.~\eqref{eq:Vbepsilon_1B}, \eqref{eq:VbThM_1B}, Eq.~\eqref{eq:Vbepsilon_1B=0}, \eqref{eq:VbTh_1B=0}, we get the full bosonic contribution to $V_{\text{eff}}$ as
	\begin{align}
		V_{b,\epsilon}^{(1,B)}+V_{b,\text{Ring},\epsilon}^{(1,B)} &= \frac{|eB|^2}{8\pi^2}\bBigg@{2.5}\{\zeta\left(-1,\frac{1}{2}+\frac{m_b^2+\Pi}{2|eB|}\right)\left(\frac{1}{\epsilon}-\gamma_E-\log\frac{2|eB|}{4\pi \tilde{\Lambda}^{2}}+1\right)\\
		&
		\hspace{4cm}	+\ \zeta^{(1,0)}\left(-1,\frac{1}{2}  +\frac{m_b^2+\Pi}{2|eB|}\right)\bBigg@{2.5}\}+\mathcal{O}(\epsilon), \\
		V^{(1,B)}_{b,\text{ThM}} + V_{b,\text{Ring},\text{ThM}}^{(1,B)} &=  \frac{|eB|}{2\pi^2}T\sum_{\ell=0}^{\infty}\int\limits^{\infty}_{0}dk_z\,\,\log\left[1-\exp\left(-\frac{\sqrt{k_z^2+(2\ell+1)|eB|+m_b^2+\Pi}}{T}\right)\right].
	\end{align}
	Since $M_{f} = g v$ with $g, v>0$, the quark contribution to the effective potential never becomes imaginary. Consequently, the resummation of quark contribution is not necessary at this point.
	\begin{figure}[!h]
		\centering
		\includegraphics[scale=0.4]{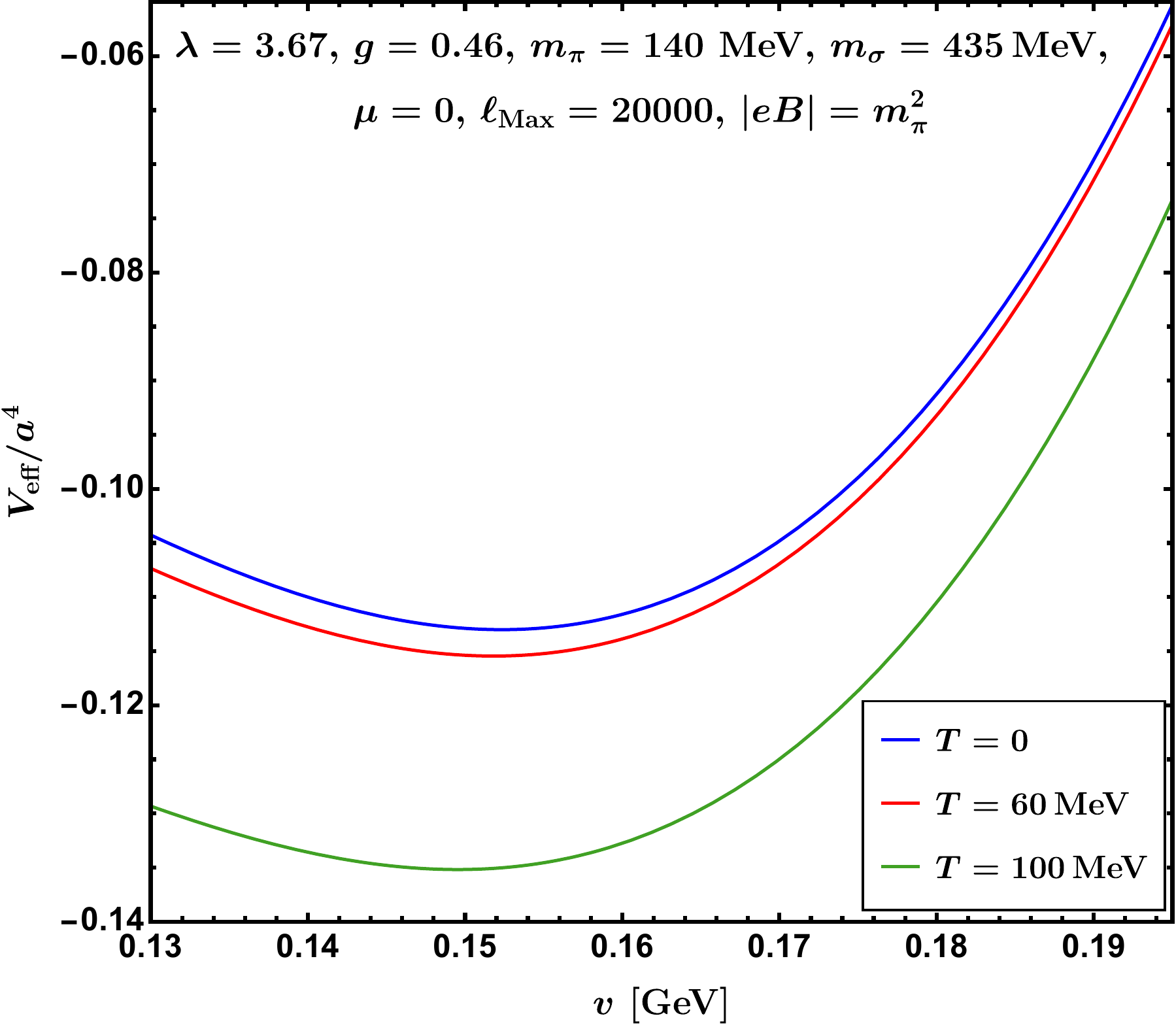}
		\includegraphics[scale=0.4]{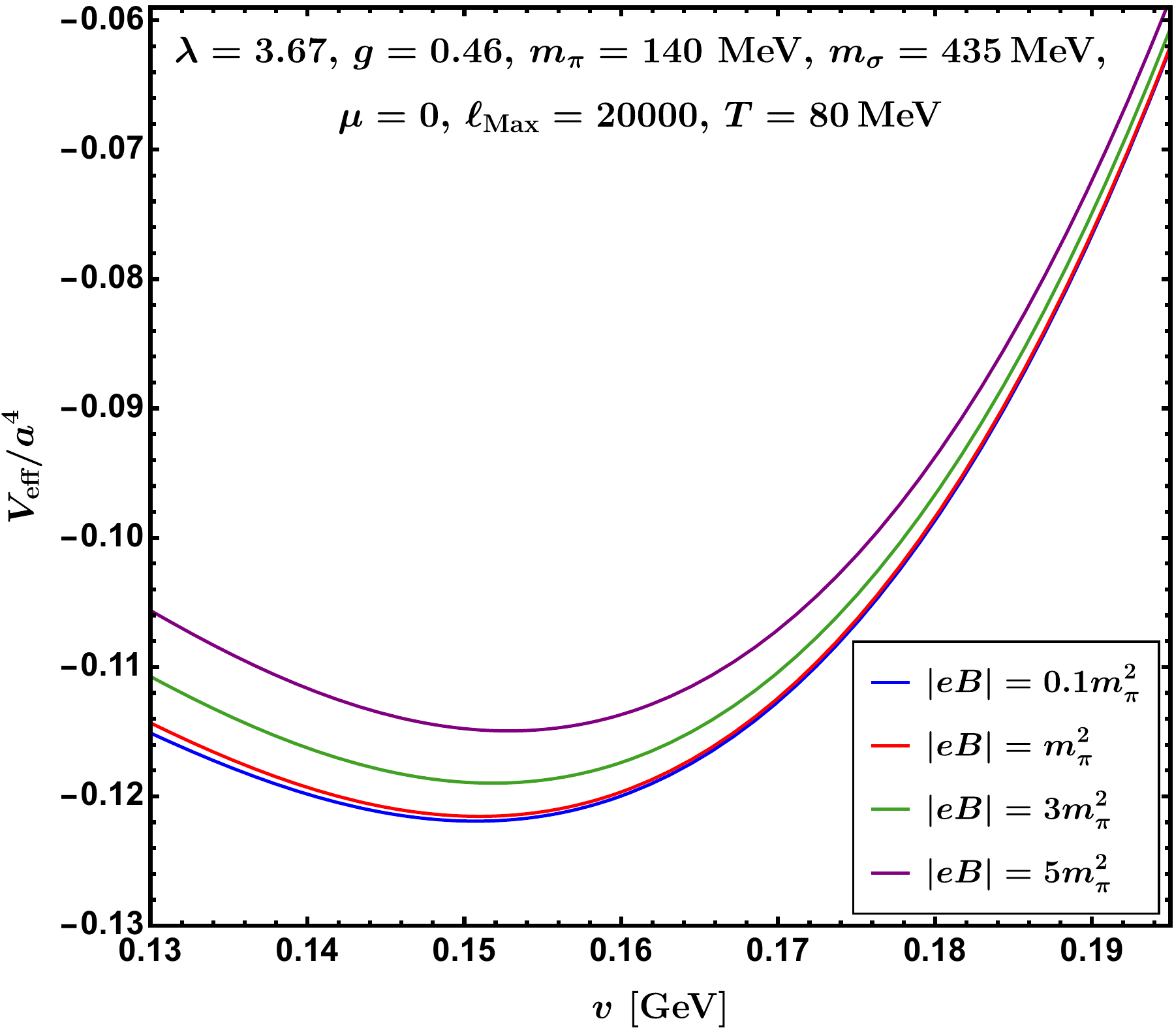}
		\label{fig:eff_pot_plots}
		\caption{[Color online] Left panel shows plot of effective potential as a function of $v$ with a fixed $eB$ with different $T$ whereas the right panel shows that with a fixed $T$ for different $eB$.}
	\end{figure}
	\section{Perpendicular Momentum Integrations}\label{sec:identities}
	In this section, we perform the general perpendicular integration  
	\begin{align}
		&\mathcal{I}^{(\alpha)}_{\ell,n}\equiv \int\frac{d^2k_{\sperp}}{(2\pi)^2}\exp\left(-\frac{2k^2_{\sperp}}{|q_{f}B|}\right)k_{\sperp}^{2\alpha}L_{\ell}\left(\frac{2k^2_{\sperp}}{|q_{f}B|}\right) L_{n}\left(\frac{2k^2_{\sperp}}{|q_{f}B|}\right),
		\label{eq:perp_intg_gen}
	\end{align} 
	where $\ell,n,\alpha$ are integers and $\ell,n,\alpha\geq 0$ and $d^2k_{\perp}\equiv dk^1dk^2=d\phi\, dk_{\sperp}\,k_{\sperp}$. After a change of variable $\xi=2k_{\sperp}^2/|q_{f}B|$, we get 
	\begin{align}
		&\mathcal{I}^{(\alpha)}_{\ell,n}=\frac{|q_{f}B|}{8\pi}\left(\frac{|q_{f}B|}{2}\right)^{\alpha}\int_{0}^{\infty}d\xi\, \xi^{\alpha}e^{-\xi}L^{(\alpha)}_{\ell}(\xi)L^{(\alpha)}_{n}(\xi) . 
	\end{align}
	The generalized Laguerre polynomial satisfies the orthogonality relation:
	\begin{equation}
		\int_{0}^{\infty}dx\, x^{\alpha}e^{-x}L^{(\alpha)}_{\ell}(x)L^{(\alpha)}_{n}(x)=\dfrac{\Gamma(n+\alpha+1)}{n!}\delta_{\ell,n}, \nn
	\end{equation}
	Thus, we get
	\begin{align}
		\mathcal{I}^{(\alpha)}_{\ell,n} &= \frac{|q_{f}B|}{8\pi}\left(\frac{|q_{f}B|}{2}\right)^{\alpha}\frac{\Gamma(\ell+\alpha +1)}{\ell !}\,\delta_{\ell,n}, \label{eq:perp_intg_result}
	\end{align}
	The two most important perpendicular integral in our context is obtained by setting $\alpha=0,1$ in \eqref{eq:perp_intg_result} as
	\begin{align}
		\int\frac{d^2k_{\sperp}}{(2\pi)^2}\exp\(-\dfrac{2k_{\sperp}^2}{|q_fB|}\)L_{\ell}\(\dfrac{2k_{\sperp}^2}{|q_fB|}\)L_{n}\(\dfrac{2k_{\sperp}^2}{|q_fB|}\) =\frac{|q_fB|}{8\pi}\delta_{\ell,n}\,\,\,\, ,\\
		\int\frac{d^2k_{\sperp}}{(2\pi)^2}k^2_{\sperp}\exp\(-\dfrac{2k_{\sperp}^2}{|q_fB|}\)L_{\ell}\(\dfrac{2k_{\sperp}^2}{|q_fB|}\)L_{n}\(\dfrac{2k_{\sperp}^2}{|q_fB|}\) =\frac{|q_fB|^2}{16\pi}(\ell+1)\delta_{\ell,n}.
	\end{align} 	  	 

\end{document}